\documentclass[aps,prx,preprint,onecolumn,citeautoscript,footinbib,superscriptaddress,
eqsecnum]{revtex4-2}
\usepackage{amsmath,amssymb}
\usepackage{graphicx}
\usepackage{dcolumn}
\usepackage{bm}
\usepackage{braket}
\usepackage{verbatim}
\usepackage[caption = false]{subfig}
\usepackage{color}
\usepackage{tcolorbox}
\usepackage{xcolor}
\usepackage{relsize}
\usepackage{amsthm}
\usepackage{enumerate}
\usepackage{physics}
\usepackage{soul,xcolor}
\usepackage[T1]{fontenc}

\usepackage[papersize={8.5in,11in}]{geometry}
\usepackage{hyperref}
\hypersetup{
    colorlinks=true,        
    linkcolor=blue,         
    citecolor=blue,         
    filecolor=blue,         
    urlcolor=blue,          
    breaklinks=true,        
    unicode=true,           
}

\usepackage{tabularray}
\UseTblrLibrary{booktabs}

\usepackage{zref-clever}
\zcsetup{cap,abbrev,nameinlink=false,comp}

\zcRefTypeSetup{equation}{name-sg = equation, name-pl = equations, name-sg-ab = eq., name-pl-ab = eqs.}
\zcRefTypeSetup{figure}{name-sg = figure, name-pl = figures, name-sg-ab = fig., name-pl-ab = figs.}
\zcRefTypeSetup{appendix}{name-sg = appendix, name-pl = appendices, name-sg-ab = app., name-pl-ab = apps.}
\zcRefTypeSetup{section}{name-sg = section, name-pl = sections, name-sg-ab = sec., name-pl-ab = secs.}
\zcRefTypeSetup{table}{name-sg = table, name-pl = tables, name-sg-ab = tab., name-pl-ab = tabs.}

\newcommand{\eqnref}[1]{\zcref{#1}}
\newcommand{\secref}[1]{\zcref{#1}}

\newcommand{\tabref}[1]{\zcref{#1}}
\newcommand{\hiddenlink}[2]{\hyperref[#1]{#2}}
\NewDocumentCommand{\figref}{ o m }{%
    \IfNoValueTF{#1}%
        {\zcref{#2}}%
        {\hiddenlink{#2}{\zcref*{#2}(#1)}}%
}

\graphicspath{{figures/}}

\usepackage{orcidlink}
\newcommand{\bi}{{\bm{i}}}
\newcommand{\bj}{{\bm{j}}}
\newcommand{\bk}{{\bm{k}}}
\newcommand{\bq}{{\bm{q}}}
\newcommand{\bQ}{{\bm{Q}}}
\newcommand{\mbf}[1]{{\bm{#1}}}

\DeclareMathOperator{\SU}{SU}
\DeclareMathOperator{\U}{U}

\newcommand{\vi}{{\bi}}
\newcommand{\vj}{{\bj}}

\newcommand{\vk}{{\bk}}
\newcommand{\vK}{{\bm{K}}}

\newcommand{\vq}{{\bm{q}}}

\newcommand{\Z}{\mathbb{Z}}

\newcommand{\mix}{\xi}

\begin{document}

\setstcolor{red}

\title{Superconducting and charge-ordered phases\texorpdfstring{\\}{ } from Dirac quantum spin liquids on the triangular lattice}

\author{Andreas Feuerpfeil\,\orcidlink{0009-0001-0436-0332}}
\affiliation{Institut für Theoretische Physik und Astrophysik, Julius-Maximilians-Universität Würzburg, Am Hubland, Campus Süd, 97074 Würzburg, Germany}
\affiliation{Würzburg–Dresden Cluster of Excellence ctd.qmat, 97074 Würzburg and 01062 Dresden, Germany}
\affiliation{Center for Computational Quantum Physics, Flatiron Institute, 162 5th Avenue, New York, NY 10010, USA}

\author{Ronny Thomale\,\orcidlink{0000-0002-3979-8836}}
\affiliation{Institut für Theoretische Physik und Astrophysik, Julius-Maximilians-Universität Würzburg, Am Hubland, Campus Süd, 97074 Würzburg, Germany}
\affiliation{Würzburg–Dresden Cluster of Excellence ctd.qmat, 97074 Würzburg and 01062 Dresden, Germany}

\author{Subir Sachdev\,\orcidlink{0000-0002-2432-7070}}
\affiliation{Department of Physics, Harvard University, Cambridge MA 02138, USA}
\affiliation{Center for Computational Quantum Physics, Flatiron Institute, 162 5th Avenue, New York, NY 10010, USA}

\author{Pietro M. Bonetti\,\orcidlink{0000-0002-7465-7043}}
\affiliation{Max Planck Institute for Solid State Research, Stuttgart, Germany}

\date{\today}

\begin{abstract}
Triangular-lattice quantum spin-liquid insulators are observed to undergo transitions to superconductivity under pressure or doping, and exhibit enhanced terahertz conductivity when driven by mid-infrared light. We present a general theoretical framework for the emergence of superconducting and charge-ordered phases from a U(1) Dirac spin liquid with fermionic spinons, as well as from its gapped $\mathbb{Z}_2$ and chiral descendants. Numerical studies have provided substantial evidence for these spin-liquid states.

The spin-liquid phase hosts fractionalized Dirac spinons coupled to an emergent gauge field, whereas the superconducting and charge-ordered phases are conventional, with neither fractionalized excitations nor emergent gauge dynamics. The transition between these phases is driven by the Higgs condensation of spinless charge-$e$ bosonic chargons (``doublons'' and ``holons''). We show that the projective symmetry group of the Dirac spinons uniquely determines the symmetry and dispersion of the chargons, allowing us to construct an effective low-energy theory near the chargon band minima. 
Gauge-invariant composites of the chargon Higgs fields provide the order parameters characterizing the phases. The resulting phase diagram contains a rich variety of ordered states, including $d+id$ superconductivity, charge-density waves, bond-density waves, and pair-density waves.
\end{abstract}

\maketitle
\newpage 
\tableofcontents

\section{Introduction}

The triangular lattice spin liquid has been of interest since the original proposal of the resonating valence bond state~\cite{fazekas1974ground,Sachdev1992Kagome,PhysRevLett.86.1881}. Preceded by phenomenological conjectures around spin liquid candidate states such as the chiral spin liquid and Gutzwiller-projected Fermi sea states~\cite{KL87,PhysRevLett.99.097202,PhysRevB.72.045105}, recent numerical work ~\cite{Yasir16,Hu-2019,Ferrari-2019,Wietek24,Sherman-2023,Drescher-2025,Yasir25,Jiang-2026} has provided significant evidence that a U(1) spin liquid state with massless Dirac fermion spinons~\cite{Wen2002,Song1} serves as a valuable `organizing principle' for understanding the low energy quantum states: The numerical results leave open the possibility that the true ground state has a small energy gap, and the likely gapped states can be obtained as descendants of the gapless U(1) spin liquid, i.e., it is possible to add small masses to the Dirac fermions and obtain a $\mathbb{Z}_2$ spin liquid~\cite{Sachdev1992Kagome,Lu16,Jiang23,Feuerpfeil_2026} and a Kalmeyer-Laughlin chiral spin liquid numerically observed for a certain parametric window of Hubbard $U$ coupling strength succeeded by Neel order at larger U~\cite{KL87,PhysRevB.91.245125,Zaletel18}.

As quantum spin liquids are in principle rather elusive in nature and susceptible to many sources of imperfections upon comparing the effective theoretical model with actual quantum materials, careful reflection is warranted as to how to employ an organizing principle based on a spin liquid mother state. This particularly concerns quantum magnets on a triangular lattice, where anisotropic spin exchange symmetry can significantly enhance ordering tendencies while geometric anisotropy towards the quasi-1D limit quickly promotes the propensity for frustrated quantum magnetism and topologically trivial quasi-1D quantum paramagnets~\cite{PhysRevB.83.024402}.
Notwithstanding related ongoing activity in other areas realizing correlated electrons on a triangular lattice such as twisted transition metal dichalcogenides or sodium cobaltate~\cite{Takada2003,10.1063/5.0077901}, recent experimental evidence for triangular lattice spin liquids has appeared in organic materials~\cite{Kanoda18,Kanoda24,matteo26} and NaYbSe$_2$~\cite{Tennant24,chinese_na}. While the precise geometric anisotropy in organic materials  is difficult to track both experimentally and from a quantitative ab initio point of view, NaYbSe$_2$ appears to exhibit no crystallographic distortion and hence can tentatively be assumed to feature an isotropic triangular lattice. Most importantly, independent of the anisotropy issue to be addressed in the future, all aforementioned materials  provide a highly suited basis for perturbing the triangular paramagnet into adjacent electronic orders. Adopting the viewpoint of a spin liquid mother state, this amounts to triggering spin liquid descendant phases allowing for a comparison between theoretical predictions and experimental evidence. Specifically, our interest here will be on the fate of such insulating tentative spin liquids when charge fluctuations are enhanced {\it e.g.\/} by applied pressure, which leads to superconductivity in these materials. Ref.~\cite{matteo26} has observed light-driven enhanced charge conductivity in the organic compound $\kappa$-(BEDT-TTF)$_2$Cu$_2$(CN)$_3$, and we shall describe here aspects of the theory of charge fluctuations applied to this compound.

Building on the analysis of Refs.~\cite{Christos23,Bonetti26} of charge fluctuations in square lattice spin liquids, we want to apply a similar logical thread to the triangular lattice. The square lattice spin liquid mother state has massless Dirac fermion spinons coupled to a SU(2) gauge field whereas we start from a U(1) spin liquid on the triangular lattice. Here, given fermionic spinons $f$, the charge fluctuations are associated with a charge $e$ (Higgs) boson $B$ that carries a fundamental charge under the emergent U(1) gauge field. 
The boson $B$ describes charge fluctuations of the spin liquid to both doubly-occupied (``doublons'') and empty (``holons'') sites.
The central idea we put forward is that the knowledge of the symmetry transformations of $f$ is sufficient to specify the symmetry transformations of $B$ from the requirement that the fusion of $f$ and $B$ be a microscopic electron, $c$. Knowing the symmetry transformations of $B$, we can then specify the Landau-type action functional (the `Higgs potential') which controls fluctuations of $B$. We will be interested in the nature of the confining/Higgs phase where $B$ is condensed
{\it i.e.\/} all the ordered phases will be conventional, and not carry any deconfined emergent gauge fields. 
To characterize these ordered phases,
we need to compute the various gauge-invariant order parameters which can be constructed as composites of the condensed $B$, and determine which of them are non-zero in each phase. In effect, the Higgs field $B$ acts as a `fractionalized order parameter'~\cite{CSS93} for all the intertwined broken symmetries in the confining phases.

Our analysis will be restricted to the half-filled case, although our main results apply also at non-zero doping, assuming that the doped charge carriers are in the form of electrons {\it i.e.\/} in a fractionalized Fermi liquid (FL*)~\cite{YaHui-ancilla1,Christos23,Boulder25}.
The main doping-induced change in the effective action for $B$ in this case is the appearance of a term with a first-order time derivative~\cite{Christos23,Bonetti26}. The Landau energy functional of $B$ does not acquire any significant changes, and so our results for the structure of the ordered phases here are not modified.

This approach to the doped case should be contrasted to recent studies of superconductivity in doped chiral spin liquids~\cite{Shi2025,Divic2025,Pichler2026,Clemens2025}, which assume that the doped charge carriers are anyons. The superconducting state obtained from the theory of the anyon gas is confining, and so the superconductor may ultimately be the same as that obtained from the FL* approach. The distinction between the FL* and anyon theories could only be in the nature of the normal state above the superconducting critical temperature.

\section{The U(1) Dirac spin liquid and its \texorpdfstring{$\mathbb{Z}_2$}{Z2} and chiral descendants}

We investigate charge orders emerging from the Dirac spin liquid (DSL) alongside its $\mathbb{Z}_2$ and chiral descendants. To this end, we formulate the fermionic spinon theory~\cite{Wen-1991} on the triangular lattice. We begin by representing the spin operators in terms of spinons $f_{\vi\alpha}$, with $\alpha=\uparrow, \downarrow$ for sites $\vi=(i_x,i_y)$~\cite{Abrikosov-1965}:
\begin{equation}
    \mbf{S}_{\vi}=\frac{1}{2}f_{\vi\alpha}^\dag \bm{\sigma}_{\alpha\beta}f_{\vi\beta}\,.
\end{equation}
Following Wen~\cite{Wen2002}, we introduce the Nambu spinor 
\begin{equation}
    \psi_{\vi}=\begin{pmatrix}
        f_{\vi\uparrow}\\
        f_{\vi\downarrow}^\dag
    \end{pmatrix}\,,
\end{equation}
which yields the Bogoliubov mean-field Hamiltonian
\begin{equation}\label{eq:bogoliubov_hamiltonian}
    H=-\sum_{\vi\vj}\psi_{\vi}^\dag u_{\vi\vj}\psi_{\vj}\,.
\end{equation}
Here, 
\begin{equation}
    u_{\vi \vj} = iu_{\vi\vj}^0\tau^0+u_{\vi\vj}^a\tau^a
\end{equation}
is the mean-field link field, with the Pauli matrices $\tau^\mu$ acting on the Nambu spinor $\psi_{\vi}$. This spinon representation is invariant under $\SU(2)$ gauge transformations $\psi_\vi\to W_\vi \psi_\vi$, with $W_\vi\in\text{SU(2)}$~\cite{Affleck1988,Dagotto-1988}. Additionally, translations and point-group symmetries may be implemented \textit{projectively}~\cite{Wen2002}. This means that the action of a symmetry operator $\mathcal{O}$ can be written as: 
\begin{equation}
\begin{split}\label{eq:psg_transformation}
    \SU(2)_\mathcal{O}: \psi_{\vi}&\rightarrow W_{\vi}^\mathcal{O}\psi_{\mathcal{O}(\vi)}\,,\\
    u_{\vi\vj}&\rightarrow W_{\vi}^{\mathcal{O}} u_{\mathcal{O}(\vi),\mathcal{O}(\vj)}(W_{\vj}^\mathcal{O})^\dagger\,,
\end{split}    
\end{equation}
where $W_{\vi}^\mathcal{O}\in\SU(2)$.

Extensive numerical investigations---including density-matrix renormalization group (DMRG), exact diagonalization, variational Monte Carlo, and dynamical studies---report an intermediate spin-liquid regime consistent with a $\U(1)$ Dirac spin liquid for the spin-$1/2$ $J_1$--$J_2$ Heisenberg antiferromagnet on the triangular lattice~\cite{Zhu-2015,Hu-2015,Yasir16,Hu-2019,Ferrari-2019,Wietek24,Sherman-2023,Drescher-2025,Yasir25,Jiang-2026,Kovalska-2026}. To address the role of charge fluctuations, we study fractionalized orders arising from the $\U(1)$ Dirac spin liquid, as well as the $\Z_2$ and chiral spin liquids that descend from it. For our mean-field ansatz, we choose~\cite{Bieri_2016,Hu_2016}:
\begin{equation}\label{eq:ansatz}
\begin{split}
    u^{}_{(i_1,i_2),(i_1,i_2+1)}&=e^{i\phi}t\tau^3\,,\\
    u^{}_{(i_1,i_2),(i_1+1,i_2)}&=(-1)^{i_2}e^{i\phi}t\tau^3\,,\\
    u^{}_{(i_1,i_2),(i_1+1,i_2+1)}&=(-1)^{i_2}e^{-i\phi}t\tau^3\,,\\
    u^{}_{(i_1,i_2),(i_1,i_2)}&=\lambda\tau^1\,.
\end{split}
\end{equation}
Here, $\hat{\mbf{e}}_1=\left(1,0\right)$, $\hat{\mbf{e}}_2=\left(-{1}/{2},{\sqrt{3}}/{2}\right)$, and $\bi=i_1\,\hat{\mbf{e}}_1+i_2\,\hat{\mbf{e}}_2$ denote the spatial coordinates.

The $\U(1)$ $0$--$\pi$ flux Dirac spin liquid is recovered by setting $\phi=\lambda=0$. Condensing $\lambda$ higgses the $\U(1)$ Dirac spin liquid into a gapped $\Z_2$ spin liquid (labeled \#20 in Ref.~\cite{Lu16}), consistent with the bosonic parton approach detailed in Ref.~\cite{Sachdev1992Kagome}. The parameter $\lambda$ corresponds to the singlet superconducting pairing term $\Delta_{\mathrm{ssc}}$ in Ref.~\cite{Lu16}.

Introducing a finite phase $\phi \not \in \frac{\pi}{3}\mathbb{Z}$ gaps the Dirac points of the DSL, and assigns the same Chern-number to the lowest band of both spinon species, $C_\uparrow=C_\downarrow=\operatorname{sgn}(\phi)$. This transition yields a gapped chiral spin liquid (CSL) characterized by a flux of $3\phi$ through the up-triangles and $\pi-3\phi$ through the down-triangles, as illustrated in \figref[b]{fig:lattice_chiral}. We investigate this state as a potential candidate for the CSL observed in DMRG calculations of the Hubbard model on the triangular lattice~\cite{Zaletel18}. The resulting state carries a non-zero scalar spin chirality $\langle\mbf{S}_\vi\cdot(\mbf{S}_\vj\times\mbf{S}_\vk)\rangle\neq0$ and realizes Kalmeyer--Laughlin topological order~\cite{KL87}.

\begin{figure}
    \centering
    \includegraphics[width=1.0\linewidth]{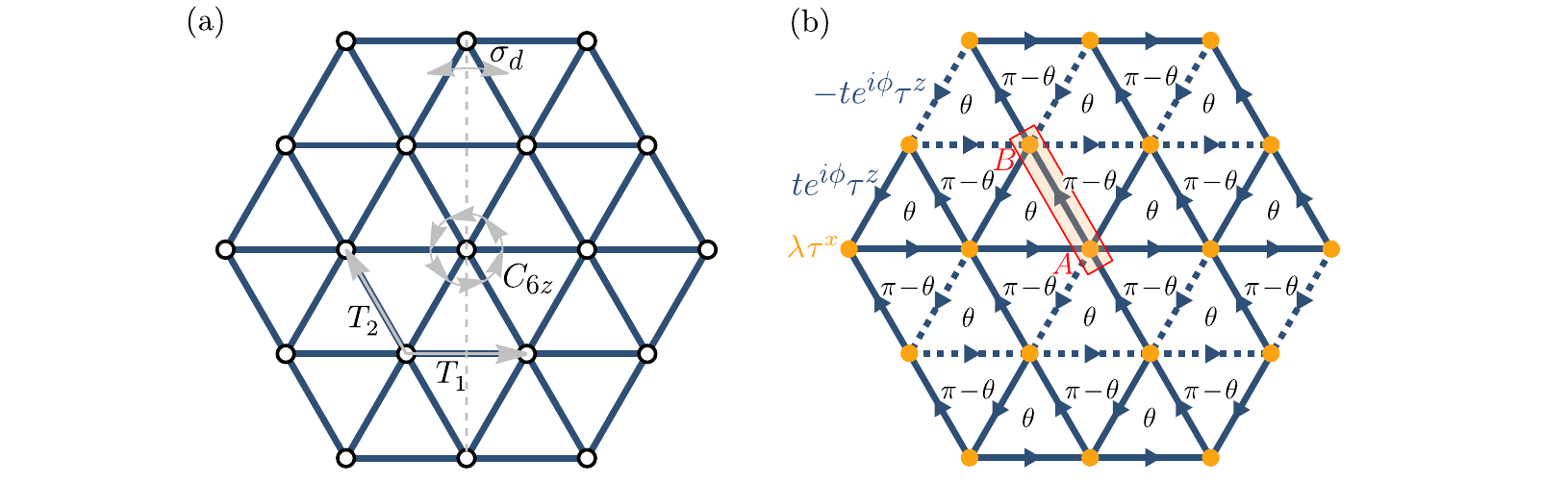}
    \caption{(a) Triangular lattice with the minimal set of space group symmetry generators. (b) Graphical illustration of the $\U(1)$ $0$--$\pi$ flux state and its daughter $\mathbb{Z}_2$ and chiral spin liquids. Realizing this ansatz requires doubling the unit cell along the $T_2$ direction, resulting in the two-site unit cell indicated by the red rectangle. The addition of $\lambda$ higgses the $\U(1)$ spin liquid into a gapped $\mathbb{Z}_2$ spin liquid. Adding a non-zero phase $\phi\neq 0$ breaks time reversal and results in a chiral spin liquid with flux $\pi-\theta$ and $\theta$ for $\theta=3\phi$. Solid (dashed) lines indicate positive (negative) sign structure and the arrow defines the orientation of $u_{\vi\vj}=u_{\vj\vi}^*$.}
    \label{fig:lattice_chiral}
\end{figure}

\begin{figure}
    \centering
    \includegraphics[width=1.0\linewidth]{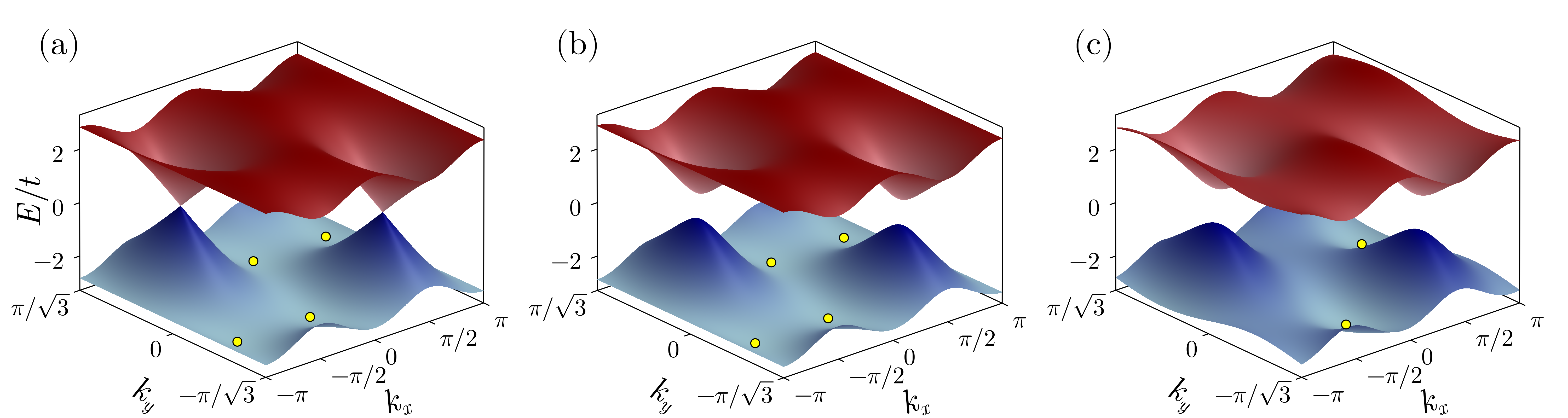}
    \caption{Spinon and chargon dispersion on the triangular lattice for the (a) $\U(1)$ Dirac spin liquid, (b) $\Z_2$ spin liquid with $\lambda=0.5t$, and (c) chiral spin liquid with $\phi = 0.2$, $\lambda=0$. The bands are two-fold degenerate. The minima, at which chargons can condense, are marked by yellow circles. For (a) and (b), the four minima correspond to $\bQ_{1},\dots,\bQ_{4}$ of \eqnref{eq:Q_minima}. For (c), there are only two minima. For $0<\phi<\pi/3$ (shown here), they are at $\bQ_{2}$ and $\bQ_{3}$; for $-\pi/3<\phi<0$, they are at $\bQ_{1}$ and $\bQ_{4}$. Both $\phi$ and $\lambda$ gap out the spectrum.}
    \label{fig:chargon_dispersion}
\end{figure}

The lattice, along with its point-group symmetries and the chosen ansatz, is depicted in \figref{fig:lattice_chiral}. The corresponding spinon dispersions, obtained after doubling the unit cell in the $\hat{\mbf{e}}_2$ direction, are shown in \figref{fig:chargon_dispersion}.

\section{Lattice theory of chargons}

Assuming that the DSL emerges as a phase of a model with charged degrees of freedom, we can represent the electron operators via an $\SU(2)$ rotor construction, for example, as
\begin{equation}
    \left(
    \begin{array}{c}
         c_{\bi,\uparrow} \\
         c^\dagger_{\bi,\downarrow}
    \end{array}
    \right) = 
    \mathcal{B}^\dagger_\bi \psi_\bi\,, \label{eq:fusion}
\end{equation}
with 
\begin{align}
    \mathcal{B}_\bi=\left(\begin{array}{cc}
    B_{\bi,+} & -B_{\bi,-}^*\\
    B_{\bi,-} & B_{\bi,+}^*
\end{array}\right)\,,
\end{align}
which encodes the bosonic degrees of freedom, which we call chargons, that carry the electron charge and pseudospin. In this representation, the spinons $\psi_\bi$ carry the physical spin but are neutral under the global electromagnetic $\U(1)$ symmetry. Instead, the physical electron charge is carried entirely by the chargons, with both $B_{\bi,+}$ and $B_{\bi,-}$ carrying the same physical charge.

This decomposition introduces an $\SU(2)$ gauge invariance $\psi_\bi\to W_\bi \psi_\bi$, $\mathcal{B}_\bi\to W_\bi \mathcal{B}_\bi$, for $W_\bi\in \SU(2)$. Under this gauge group, the chargon components form a fundamental doublet $B_\bi=(B_{\bi,+},B_{\bi,-})^\top$. The DSL hoppings higgs this $\SU(2)$ gauge group down to $\U(1)$, which causes $B_{\bi,+}$ and $B_{\bi,-}$ to carry opposite emergent $\U(1)$ gauge charges. Because the fusion of the chargons and the spinons yields an electron in \eqnref{eq:fusion}, and the electron does not experience the emergent gauge field, the effective action of the chargons must take a form similar to that of the spinons~\cite{Christos23}; at quadratic order, the action is:
\begin{equation}
    \mathcal{S}_B = \int_0^\beta\dd\tau\left[\sum_{\bi} \left((D_\tau B_{\bi})^\dagger\,D_\tau B_{\bi}+r B^\dagger_{\bi} B_{\bi}\right)+\sum_{\substack{\bi\bj}}B^\dagger_{\bi}u_{\vi\vj} B_{\bj}\right]\,, \label{eq:chargonactionU1}
\end{equation}
where $r$ is a mass parameter for the chargons.

In \tabref{tab: PSG lattice}, we list the projective symmetry group (PSG) transformations of the spinon Nambu spinor $\psi_\bi$ and the chargons $B_{\bi,a}$ for the $\U(1)$ and $\Z_2$ spin liquids. These include translations $T_1$ and $T_2$ along the $\hat{\mbf{e}}_1$ and $\hat{\mbf{e}}_2$ directions, time reversal ($\mathcal{T}$), and two representative operations of the triangular lattice point group ($D_6$): reflection across the $y$-axis ($\sigma_d$) and $\pi/3$ rotation about the $z$-axis ($C_{6z}$). Crucially, $C_{6z}$ rotations introduce a $\tau^1$ matrix that flips the $\SU(2)$ gauge index of the chargon field. This necessitates embedding the chargon theory within an $\SU(2)$ gauge structure, even though the low-energy gauge group is strictly $\U(1)$ (or $\mathbb{Z}_2$). The functions or matrices multiplying $\psi$ and $B$ in \tabref{tab: PSG lattice} correspond to the gauge transformation matrices $W^\mathcal{O}_\bi$ of \eqnref{eq:psg_transformation}.

The CSL ansatz, characterized by $\phi \neq 0$ and $\lambda = 0$, explicitly breaks time reversal ($\mathcal{T}$) and the flux-reversing spatial symmetries---specifically, reflection ($\sigma_d$) and six-fold rotation ($C_{6z}$).

The physical origin of this symmetry breaking lies in the non-zero complex phase $e^{i\phi}$ on the lattice bonds, which generates a net emergent gauge flux through the elementary triangles. Time reversal applies complex conjugation ($\phi \rightarrow -\phi$), while the spatial operations $\sigma_d$ and $C_{6z}$ map ``up'' triangles to ``down'' triangles, effectively reversing the geometric orientation of the flux. Consequently, the ansatz is not invariant under these individual transformations. However, their composite operations ($C_{6z}\mathcal{T}$ and $\sigma_d\mathcal{T}$) compensate for the flux reversal and remain unbroken, thereby defining the surviving PSG of the chiral phase.

\begin{table}[b!]
    \centering
    \begin{tblr}{
        colspec = {c  X[c] X[c] X[c]},
        cells = {m},
        rowsep = {3pt}, 
        colsep = {0pt},
    }
    \toprule
    Symmetry  & $(i_1,i_2) $ & $\psi_\sigma$ & $B_a$ \\ \midrule
         $T_1$     & $(i_1+1,i_2)$ & $\psi_\sigma$ & $B_a$ \\

         $T_2$     & $(i_1,i_2+1)$ & $(-1)^{i_1}\psi_\sigma$ & $(-1)^{i_1}B_a$ \\

         $\sigma_d$ & $(-i_1+i_2,i_2)$ & $(-1)^{\frac{i_2(i_2-1)}{2}}\psi_{\sigma}$ & $(-1)^{\frac{i_2(i_2-1)}{2}}B_a$ \\

         $C_{6z}$  & $(i_1-i_2,i_1)$ & $(-1)^{i_2(i_1+1)+\frac{i_1(i_1+1)}{2}}(i\tau^1_{\sigma\sigma'})\psi_{\sigma'} $ & $(-1)^{i_2(i_1+1)+\frac{i_1(i_1+1)}{2}}(i\tau^1_{ab})B_b$\\

         $\mathcal{T}$ & $(i_1,i_2)$ & $i\tau^2_{\sigma\sigma'}\psi^\dagger_{\sigma'}$ & $B_a$\\\bottomrule

    \end{tblr}
    \caption{Projective transformation properties of spinons (third column) and chargons (fourth column) for the $\U(1)$ and $\Z_2$ spin liquids. In the first column, we also show how the transformations act on the lattice site coordinates, $\bi=i_1\hat{\mbf{e}}_1+i_2\hat{\mbf{e}}_2$. For the chiral spin liquid ($\phi\neq0$), $\mathcal{T}$, $\sigma_d$ and $C_{6z}$ are broken. Only $T_1$, $T_2$, $C_{6z}\mathcal{T}$ and $\sigma_d\mathcal{T}$ are preserved.}
    \label{tab: PSG lattice}
\end{table}

Our chosen PSG in \tabref{tab: PSG lattice} is consistent with the $\U(1)$ and $\Z_2$ PSGs detailed in Appendix~A of Ref.~\cite{Feuerpfeil_2026}; therefore, the $\U(1)$-breaking term $\lambda$ transforms trivially under the projective symmetry transformations.

We now construct the low-energy continuum field theory for the chargons to identify their condensation patterns. Whenever the chargons condense, the U(1) gauge group of the DSL will be fully higgsed, but additional physical symmetries can be broken, such as translations, point group symmetries, or even the U(1) charge symmetry.

To render the mean-field ansatz translationally invariant, we double the unit cell in the $\hat{\mbf{e}}_2$ direction as shown in \figref[b]{fig:lattice_chiral}.
We can then Fourier transform the mean-field Hamiltonian into momentum space:
\begin{equation}\label{eq:Hmf_sublattice}
\begin{split}
    H^{\mathrm{sub}}(\vk)=2t \tau^3 \left[\cos(k_x+\phi)\tilde{\rho}^3+\cos\left(\frac{k_x-\sqrt{3}k_y}{2}-\phi\right)\tilde{\rho}^1-\sin\left(\frac{k_x+\sqrt{3}k_y}{2}-\phi\right)\tilde{\rho}^2\right]+\lambda \tau^1 \tilde{\rho}^0\,,
\end{split}
\end{equation}
where $\tilde{\rho}^a$ are the Pauli matrices acting on the sublattice space. Changing to a ``momentum space basis`` $(\vk,\vk+\bQ_{M_1})$ with $\bQ_{M_1}=\frac{2\pi}{\sqrt{3}}(0,1)^\top$, the Hamiltonian reads
\begin{equation}\label{eq:Hmf_momentum}
\begin{split}
    H(\vk)=2t \tau^3 \left[\cos(k_x+\phi)\rho^1+\cos\left(\frac{k_x-\sqrt{3}k_y}{2}-\phi\right)\rho^3+\sin\left(\frac{k_x+\sqrt{3}k_y}{2}-\phi\right)\rho^2\right]+\lambda \tau^1 \rho^0\,.
\end{split}
\end{equation}
Note that this corresponds to the unitary transformation $H^{\mathrm{mom}}=PH^{\mathrm{sub}}P$ with $P=\frac{1}{\sqrt{2}}\left(\begin{smallmatrix}
    1 & 1 \\ 1 & -1
\end{smallmatrix}\right)$. Here, $\rho^a$ acts on the momentum index of $(\vk,\vk+\bQ_{M_1})$.

The dispersion of the Hamiltonian is given by
\begin{equation}
    \pm\epsilon_\vk=\pm\sqrt{4t^2\left(\cos^2(k_x+\phi)+\cos^2\left(\frac{k_x-\sqrt{3}k_y}{2}-\phi\right)+\sin^2\left(\frac{k_x+\sqrt{3}k_y}{2}-\phi\right)\right)+\lambda^2}\,.
\end{equation}

In the reduced Brillouin zone, defined by  $\left\{(k_x,k_y) : k_x\in(-\pi,\pi] \text{ and } k_y\in\left(-\pi/\sqrt{3},\pi/\sqrt{3}\right]\right\}$, the band structure for $\phi=0$ has four minima located at 
\begin{equation}\label{eq:Q_minima}
    \begin{aligned}
        & \bQ_1=\left(-\frac{5\pi}{6},-\frac{\pi}{2\sqrt{3}}\right)\,,\quad
        &&\bQ_2=\left(\frac{5\pi}{6},\frac{\pi}{2\sqrt{3}}\right)\,,\\
        & \bQ_3=\left(-\frac{\pi}{6},-\frac{\pi}{2\sqrt{3}}\right)\,,
        &&\bQ_4=\left(\frac{\pi}{6},\frac{\pi}{2\sqrt{3}}\right)\,.
    \end{aligned}
\end{equation}
At each of the four valleys, the Hilbert space $\mathcal{H}=\mathcal{H}_\tau\otimes\mathcal{H}_\rho$
($\dim\mathcal{H}=4$) is spanned by the chargon index $|a\rangle=|+\rangle,|-\rangle$ (Pauli matrices $\tau^i$) and the
momentum index $|m\rangle\in\{|\bQ_\eta\rangle,|\bQ_\eta+\bQ_M\rangle\}$ (Pauli matrices $\rho^i$). We can thus expand the two-fold degenerate spectrum around the four minima in terms of low-energy states
\begin{equation}
    B_{\vi}=\sum_{\eta=1}^4\sum_{\alpha=1}^2 B_{\eta,\alpha}\ket{\phi_{\eta,\alpha}}\,,
\end{equation}
where $\ket{\phi_{\eta,\alpha}}$ is a vector in $\mathcal{H}$. Here, $\ket{\phi_{\eta,\alpha}}$ depends on the lattice coordinate $\vi$ due to the fact that $\langle \mathbf{R}_\vi|\bq\rangle= e^{i\bq\cdot \mathbf{R}_\vi}$.

\subsection{U(1) and \texorpdfstring{$\Z_2$}{Z2} spin liquids}
We first consider the case where $\phi=0$. For $\lambda=0$, the $a$ sectors decouple, and the spectra are inverted by $\tau^3_{aa}$. An eigenbasis for the Hamiltonian is given by
\begin{equation}\label{eq:higgs_eigenbasis_hopping}
    \ket{+} \otimes \ket{\chi_1^\eta}\,,\quad \ket{-} \otimes \ket{\chi_2^\eta}\,,\quad \ket{+} \otimes \ket{\chi_2^\eta}\,,\quad \ket{-}\otimes \ket{\chi_1^\eta}
\end{equation}
with
\begin{equation}
\begin{split}
    \ket{\chi_1^\eta}&=-e^{i\varphi_\eta}\sin\left(\frac{\gamma}{2}\right)|\bQ_\eta\rangle + e^{-i\varphi_\eta}\cos\left(\frac{\gamma}{2}\right)|\bQ_\eta+\bQ_{M_1}\rangle\,, \\ \ket{\chi_2^\eta}&=e^{i\varphi_\eta}\cos\left(\frac{\gamma}{2}\right)|\bQ_\eta\rangle + e^{-i\varphi_\eta}\sin\left(\frac{\gamma}{2}\right)|\bQ_\eta+\bQ_{M_1}\rangle\,,
\end{split}
\end{equation}
where $\gamma=\arctan(\sqrt{2})$ and 
\begin{equation}
    \varphi_\eta=\left(\frac{3\pi}{8},-\frac{3\pi}{8},\frac{\pi}{8},-\frac{\pi}{8}\right)_\eta\,.
\end{equation}
The states possess energies $\{-\epsilon,-\epsilon,\epsilon,\epsilon\}$ with $\epsilon=3t$. Due to the orthogonality of $\ket{\chi_1^\eta}$ and $\ket{\chi_2^\eta}$, the pairing term $\lambda \tau^1$ only couples $\ket{+}\otimes \ket{\chi_m}$ and $\ket{-}\otimes \ket{\chi_m}$. Consequently, the effective Hamiltonian takes the form
\begin{equation}
    H=\begin{pmatrix}
        -\epsilon & 0 & 0 & \lambda \\
         0& -\epsilon & \lambda & 0\\
         0& \lambda& \epsilon &0 \\
         \lambda& 0& 0& \epsilon 
    \end{pmatrix}\,.
\end{equation}

Defining the mixing angle $\mix=\arctan(\lambda/\epsilon)$, the two low-energy degrees of freedom at each valley are given by 
\begin{equation}
\begin{split}
    \ket{\phi_{\eta,1}}&=\left(\cos\left(\frac{\mix}{2}\right)\ket{+}-\sin\left(\frac{\mix}{2}\right)\ket{-}\right)\otimes\ket{\chi_1^\eta}\,,\\
    \ket{\phi_{\eta,2}}&=\left(- \sin\left(\frac{\mix}{2}\right)\ket{+}+\cos\left(\frac{\mix}{2}\right)\ket{-}\right)\otimes \ket{\chi_2^\eta}\,.
\end{split}
\end{equation}
We introduce the Pauli matrices $\kappa^a$ to act on the $\alpha$ index of $\ket{\phi_{\eta,\alpha}}$. For the $\U(1)$ DSL, $\mix=0$ and $\alpha$ coincides with the $\SU(2)$ gauge index $a$ of the chargons $B_{\vi,a}$. The PSG transformations of the low-energy fields are summarized in \tabref{tab: PSG continuum}.

\begin{table}[t!]
    \centering
    \begin{tblr}{
        colspec = {c  X[c] X[c] X[c] X[c]},
        cells = {m},
        rowsep = {3pt}, 
        colsep = {0pt},
    }
    \toprule
    Symmetry & $B_{1}$ & $B_{2}$ & $B_{3}$ & $B_{4}$ \\ \midrule
         $T_1$     
         & $e^{i\frac{5\pi}{6}} B_{1}$ 
         & $e^{-i\frac{5\pi}{6}} B_{2}$
         & $e^{i\frac{\pi}{6}} B_{3}$&
         $e^{-i\frac{\pi}{6}} B_{4}$\\

         $T_2$     
         & $e^{i\frac{\pi}{3}}B_4$ 
         &$e^{-i\frac{\pi}{3}}B_3$
         &$e^{i\frac{2\pi}{3}}B_2$ 
         & $e^{-i\frac{2\pi}{3}} B_1$\\

         $\sigma_d$ 
         & $e^{i\frac{\pi}{4}}\,e^{i\frac{2\pi}{3}\kappa^3} B_2$ 
         &$e^{-i\frac{\pi}{4}}\,e^{-i\frac{2\pi}{3}\kappa^3} B_1$ 
         & $e^{i\frac{\pi}{4}}\,e^{i\frac{\pi}{3}\kappa^3} B_4$ 
         & $e^{-i\frac{\pi}{4}}\,e^{-i\frac{\pi}{3}\kappa^3} B_3$\\

         $C_{6z}$  & $ie^{i\frac{\pi}{4}}e^{-i\frac{\pi}{6}\kappa^3}\!\!\kappa^1\!\!\left(\frac{B_2 +i B_3}{\sqrt{2}}\right)$
         & $ie^{-i\frac{\pi}{4}}e^{i\frac{\pi}{6}\kappa^3}\!\!\kappa^1\!\!\left(\frac{B_1 -i B_4}{\sqrt{2}}\right)$
         & $ie^{-i\frac{\pi}{4}}e^{i\frac{\pi}{6}\kappa^3}\!\!\kappa^1\!\!\left(\frac{B_1 +i B_4}{\sqrt{2}}\right)$
         &$ie^{i\frac{\pi}{4}}e^{-i\frac{\pi}{6}\kappa^3}\!\!\kappa^1\!\!\left(\frac{B_2 -i B_3}{\sqrt{2}}\right)$\\
         
         $\mathcal{T}$ & $B_2$& $B_1$ & $B_4$ & $B_3$\\\bottomrule
    \end{tblr}
    \caption{Projective symmetry transformations of the low-energy fields $B_{\eta,\alpha}$ for the $\U(1)$ and $\Z_2$ spin liquids. For the chiral spin liquid, only $B_{1}$ and $B_{4}$ (or $B_2$ and $B_3$) are global minima. Correspondingly, only the composite symmetries $\sigma_d\mathcal{T}$ or $C_{6z}\mathcal{T}$ are conserved, rather than $\sigma_d$, $\mathcal{T}$, and $C_{6z}$ individually.}
    \label{tab: PSG continuum}
\end{table}

We can then expand the chargon field $B_{\vi,a}$ in terms of the low-energy modes $B_{\eta,\alpha}$:
\begin{equation}\label{eq: continuum expansion Bia}
    B_{\vi,a}=\sum_{\eta=1}^4\sum_{\alpha=1}^2B_{\eta,\alpha} \left[\cos\left(\frac{\mix}{2}\right)\delta_{a,\alpha}-\sin\left(\frac{\mix}{2}\right)\sigma^1_{a,\alpha}\right]\left(u_{\eta,\alpha}e^{i\bQ_\eta\cdot\vi}+v_{\eta,\alpha}e^{i(\bQ_\eta+\bQ_{M_1})\cdot\vi}\right)
\end{equation}
with 
\begin{equation}
    \begin{aligned}
        &u_{\eta,\alpha} = e^{i\varphi_\eta}\left(
        \begin{array}{c}
             -\sin\left(\frac{\gamma}{2}\right) \\
             \cos\left(\frac{\gamma}{2}\right)
        \end{array}
        \right)_\alpha\,, \hskip 1cm 
        v_{\eta,\alpha} = e^{-i\varphi_\eta}\left(
        \begin{array}{c}
             \cos\left(\frac{\gamma}{2}\right) \\
             \sin\left(\frac{\gamma}{2}\right)
        \end{array}
        \right)_\alpha\,.
    \end{aligned}
\end{equation}

\subsection{Chiral spin liquid}\label{sec:chargon_minima:CSL}
For non-zero flux $\phi \neq 0$, the four-fold degeneracy of the Dirac nodes is lifted. The emergent mass term shifts the energies of the valleys, leaving only two degenerate global minima at the original momenta: $\bQ_{2}$ and $\bQ_{3}$ for $0 < \phi < \pi/3$, and $\bQ_{1}$ and $\bQ_{4}$ for $-\pi/3 < \phi < 0$ (see \figref[c]{fig:chargon_dispersion}.

At these global minima, the low-energy physics simplifies significantly. By applying trigonometric addition formulas to the lattice momentum arguments, the exact mean-field Hamiltonian cleanly separates into a $\phi$-independent kinetic term rescaled by $\cos\phi$, and a mass term proportional to $\sin\phi$. For small momentum fluctuations $\vq$ around the unperturbed nodes $\vk = \bQ_\eta + \vq$, the mass term evaluates to a constant energy shift at zeroth order. The effective low-energy kinetic Hamiltonian is therefore simply the unperturbed Dirac spin liquid Hamiltonian $H(\vq)$ scaled by $\cos\phi$.

Because the matrix structure of the low-energy Hamiltonian is only modified by a global scalar factor, its eigenvectors---and therefore the low-energy chargon modes $B_{\eta,a}$---remain completely unaffected. Consequently, their symmetry transformations are unchanged, and all order parameters for the CSL can be obtained by simply dropping the terms containing the two higher-energy chargon valleys from the $\U(1)$ Dirac spin liquid.

The non-zero Chern numbers of the spinon bands imply that the U(1) gauge field of the CSL has a Chern-Simons term. While this term will modify the critical properties of the $B$ condensation transition, it does not modify the analysis at the mean-field level of the present paper.

\section{Symmetry-breaking patterns}

To identify the different symmetry-breaking patterns arising from the condensation of $B_{\vi,a}$ at one or more of the four minima, we construct SU(2) gauge-invariant bilinears:
\begin{subequations}\label{eq: gauge inv bilinears}
    \begin{align}
        &\rho_\bi = B^\dagger_{\bi} B_{\bi}\,,\\
        &b_{\bi\bj} = B^\dagger_{\bi} u_{\bi\bj}\,B_{\bj}\,,\\
        &Q_{\bi\bj}=\frac{b_{\bi\bj}+b_{\bj\bi}}{2}\,,\\
        &J_{\bi\bj}=\frac{b_{\bi\bj}-b_{\bj\bi}}{2i}\,,\\
        &\Delta_{\bi\bj} = B_{\bi}^\top (-i\tau^2)u_{\bi\bj} B_\bj\,.
    \end{align}
\end{subequations}

A non-uniform value of $\rho_\bi$ or $Q_{\bi\bj}$ indicates the formation of charge density wave (CDW) or bond density wave (BDW) order, where $\rho_\bi$ represents the onsite density and $Q_{\bi\bj}$ the bond density. $J_{\vi\vj}$ is a current order parameter that spontaneously breaks time-reversal symmetry. Finally, $\Delta_{\bi\bj}$ is a pairing order parameter, signaling the spontaneous breaking of the global U(1) charge symmetry. 

\subsection{Symmetry-breaking patterns of the U(1) spin liquid}\label{sec:U1_orders}
For the Dirac spin liquid, the form of the low-energy modes simplifies because $\mix=0$. In the following, we tabulate all possible bilinears that can be formed by condensing chargons at the four minima defined in \eqnref{eq:Q_minima}.

\subsubsection{CDW and BDW orders}
\begin{figure}[t!]
    \centering
    \includegraphics[width=0.75\textwidth]{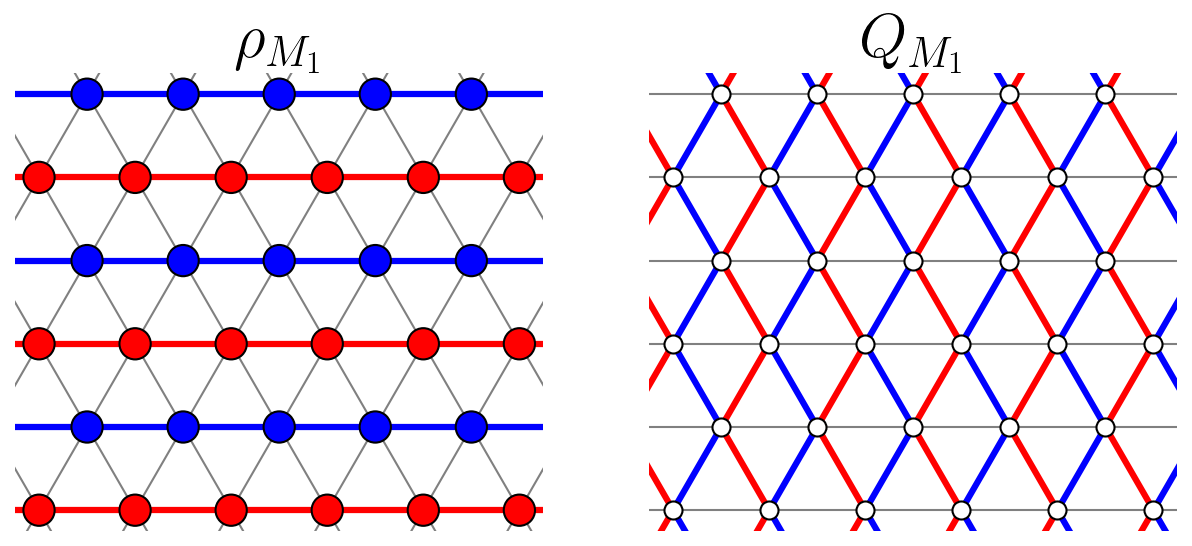}
    \caption{Charge and bond densities induced by the condensation of the $\rho_{M_1}$ (left) and $Q_{M_1}$ (right) order parameters. Red (blue) thick bonds and sites represent a higher charge or bond density, whereas thin gray bonds and white sites indicate no deviation in density.}
    \label{fig: rhoM1 and QM1}
\end{figure}
\begin{figure}[t!]
    \centering
        \includegraphics[width=0.75\textwidth]{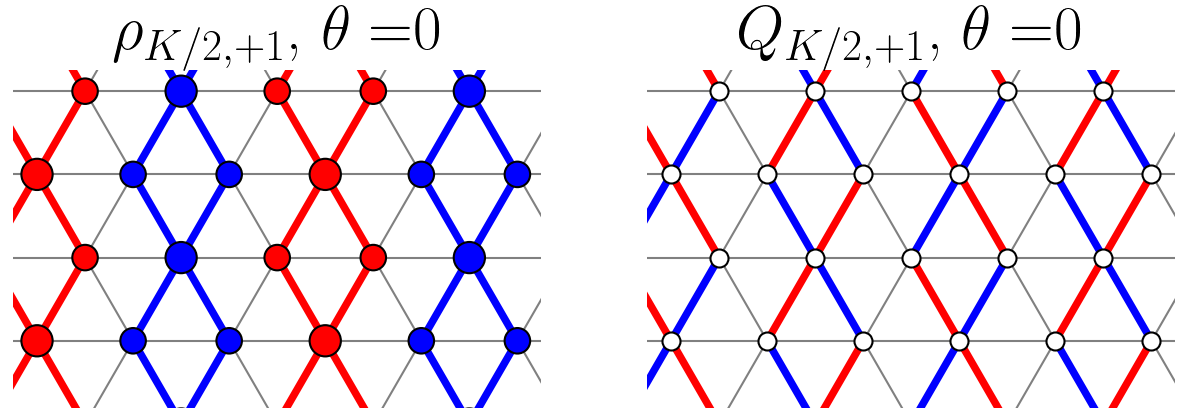}
    \caption{Charge and bond densities induced by the condensation of the $\rho_{K/2,1}$ (left) and $Q_{K/2,1}$ (right) order parameters. Red (blue) thick bonds and sites represent a higher charge or bond density, whereas thin gray bonds and sites indicate no deviation in density. 
    }
    \label{fig: rhoK21 and QK21}
\end{figure}
By inserting \eqnref{eq: continuum expansion Bia} into \eqnref{eq: gauge inv bilinears}, straightforward algebra reveals that CDW and BDW orders can form at the three $M$-points ($\bQ_{M_1}=\left(0,2\pi/\sqrt{3}\right)$, $\bQ_{M_2}=\left(-\pi,\pi/\sqrt{3}\right)$, $\bQ_{M_3}=\left(\pi,\pi/\sqrt{3}\right)$), yielding period-2 CDWs. Condensation can also occur at the wave vectors $\bQ_{K/2,1}=\left(2\pi/3,0\right)$, $\bQ_{K/2,2}=\left(\pi/3,\pi/\sqrt{3}\right)$, and $\bQ_{K/2,3}=\left(-\pi/3,\pi/\sqrt{3}\right)$, producing period-6 CDWs. The order parameters at these various wave vectors can be expressed in terms of bilinears of the continuous fields $B_{\eta,a}$ as:
\begin{subequations}
    \begin{align}
        &\rho_{M_1}=\left(B^\dagger_{1}\kappa^3B_{1}+B^\dagger_{2}\kappa^3B_{2}-B^\dagger_{3}\kappa^3B_{3}-B^\dagger_{4}\kappa^3B_{4}\right)\,,\\
        &\rho_{M_2}=-\left(B^\dagger_{1}\kappa^3B_{4}+B^\dagger_{4}\kappa^3B_{1}+B^\dagger_{3}\kappa^3B_{2}+B^\dagger_{2}\kappa^3B_{3}\right)\,,\\
        &\rho_{M_3}=i\left(B^\dagger_{1}\kappa^3B_{4}-B^\dagger_{4}\kappa^3B_{1}+B^\dagger_{3}\kappa^3B_{2}-B^\dagger_{2}\kappa^3B_{3}\right)\,,\\
        \nonumber\\
        &\rho_{K/2,1}=B_3^\dagger\,e^{-i\frac{\pi}{6}\kappa^3}\,B_1+B_2^\dagger\,e^{-i\frac{\pi}{6}\kappa^3}\,B_4\,,\\
        &\rho_{K/2,2}=B_1^\dagger\,e^{i\frac{5\pi}{6}\kappa^3}\,B_2+B_4^\dagger\,e^{-i\frac{\pi}{6}\kappa^3}\,B_3\,,\\
        &\rho_{K/2,3}=B^\dagger_2\,(-\kappa^3)\,B_1+B^\dagger_3\,(-\kappa^3)\,B_4\,.
    \end{align}
\end{subequations}
Additionally, we find bond density waves with a vanishing modulation of the onsite charge density. They are given by:
\begin{subequations}
    \begin{align}
        &Q_{M_1}=\left(B^\dagger_{1}B_{1}+B^\dagger_{2}B_{2}-B^\dagger_{3}B_{3}-B^\dagger_{4}B_{4}\right)\,,\\
        &Q_{M_2}=-\left(B^\dagger_{1}B_{4}+B^\dagger_{4}B_{1}+B^\dagger_{3}B_{2}+B^\dagger_{2}B_{3}\right)\,,\\
        &Q_{M_3}=i\left(B^\dagger_{1}B_{4}-B^\dagger_{4}B_{1}+B^\dagger_{3}B_{2}-B^\dagger_{2}B_{3}\right)\,,\\
        \nonumber\\
        &Q_{K/2,+1}=B_3^\dagger\,e^{-i\frac{2\pi}{3}\kappa^3}\,B_1+B_2^\dagger\,e^{-i\frac{2\pi}{3}\kappa^3}\,B_4\,,\\
        &Q_{K/2,+2}=B_1^\dagger\,e^{i\frac{\pi}{3}\kappa^3}\,B_2+B_4^\dagger\,e^{-i\frac{2\pi}{3}\kappa^3}\,B_3\,,\\
        &Q_{K/2,+3}=i\left(B^\dagger_2\,B_1+B^\dagger_3\,B_4\right)\,.
    \end{align}
\end{subequations}
In \figref{fig: rhoM1 and QM1}, we display the site and bond densities of a state where $B_{\eta,\alpha}$ condense such that the bilinears $\rho_{M_1}$ or $Q_{M_1}$ are non-vanishing. Similarly, \figref{fig: rhoK21 and QK21} details the same for $\rho_{K/2,1}$ and $Q_{K/2,1}$. 

\begin{table}
    \centering
    \begin{tblr}{
        colspec = {c  X[c] X[c] X[c] X[c] X[c] X[c]},
        cells = {m},
        rowsep = {3pt}, 
        colsep = {0pt},
    }
    \toprule
    Symmetry & $\rho_{M_1}$ & $\rho_{M_2}$ & $\rho_{M_3}$ 
        & $\rho_{K/2,1}$ & $\rho_{K/2,2}$ & $\rho_{K/2,3}$ \\ \midrule

         $T_1$ & $\rho_{M_1}$ & $-\rho_{M_2}$ & $-\rho_{M_3}$ 
         & $e^{i\frac{2\pi}{3}}\rho_{K/2,1}$ & $e^{i\frac{\pi}{3}}\rho_{K/2,2}$ &  $e^{-i\frac{\pi}{3}}\rho_{K/2,3}$ \\

         $T_2$ & $-\rho_{M_1}$ &  $-\rho_{M_2}$ & $\rho_{M_3}$ 
         &  $e^{-i\frac{\pi}{3}}\rho_{K/2,1}$ & $e^{i\frac{\pi}{3}}\rho_{K/2,2}$ & $e^{i\frac{2\pi}{3}}\rho_{K/2,3}$\\

         $\sigma_d$ & $\rho_{M_1}$ & $\rho_{M_3}$ & $\rho_{M_2}$ 
         & $\rho_{K/2,1}^*$ & $\rho_{K/2,3}$ & $\rho_{K/2,2}$\\
         
         $C_{6z}$  & $\rho_{M_3}$ &$\rho_{M_1}$ & $\rho_{M_2}$ 
         & $\rho_{K/2,3}^*$ & $\rho_{K/2,1}$ & $\rho_{K/2,2}$\\

         $\mathcal{T}$ &$\rho_{M_1}$ &$\rho_{M_2}$ & $\rho_{M_3}$ 
         & $\rho_{K/2,1}^*$ & $\rho_{K/2,2}^*$ & $\rho_{K/2,3}^*$ \\\bottomrule
    \end{tblr}
    \caption{Transformation properties of the CDW order parameters.}
    \label{tab: transf. rho}
\end{table}

\begin{table}[b]
    \centering
    \begin{tblr}{
        colspec = {c  X[c] X[c] X[c] X[c] X[c] X[c]},
        cells = {m},
        rowsep = {3pt}, 
        colsep = {0pt},
    }
    \toprule
    Symmetry  & $Q_{M_1}$ & $Q_{M_2}$ & $Q_{M_3}$ 
        & $Q_{K/2,1}$ & $Q_{K/2,2}$ & $Q_{K/2,3}$ \\ \midrule

         $T_1$ & $Q_{M_1}$ & $-Q_{M_2}$ & $-Q_{M_3}$ 
         & $e^{i\frac{2\pi}{3}}Q_{K/2,1}$ & $e^{i\frac{\pi}{3}}Q_{K/2,2}$ &  $e^{-i\frac{\pi}{3}}Q_{K/2,3}$ \\

         $T_2$ & $-Q_{M_1}$ &  $-Q_{M_2}$ & $Q_{M_3}$ 
         &  $e^{-i\frac{\pi}{3}}Q_{K/2,1}$ & $e^{i\frac{\pi}{3}}Q_{K/2,2}$ & $e^{i\frac{2\pi}{3}}Q_{K/2,3}$\\

         $\sigma_d$ & $Q_{M_1}$ & $Q_{M_3}$ & $Q_{M_2}$ 
         & $-Q_{K/2,1}^*$ & $Q_{K/2,3}$ & $Q_{K/2,2}$\\

         $C_{6z}$  & $-Q_{M_3}$ &$-Q_{M_1}$ & $-Q_{M_2}$ 
         & $Q_{K/2,3}^*$ & $Q_{K/2,1}$ & $Q_{K/2,2}$\\

         $\mathcal{T}$ &$Q_{M_1}$ &$Q_{M_2}$ & $Q_{M_3}$ & $Q_{K/2,1}^*$ & $Q_{K/2,2}^*$ & $Q_{K/2,3}^*$ \\\bottomrule

    \end{tblr}
    \caption{Transformation properties of the BDW order parameters.}
    \label{tab: transf BDW}
\end{table}

We note that because $\bQ_{K/2,n}$ (for $n=1,2,3$) is not equal to $-\bQ_{K/2,n}$ modulo a reciprocal lattice vector, the bilinears $\rho_{K/2,n}$ take complex values, unlike $\rho_{M_n}$. This implies that different choices of the complex phase yield distinct spatial patterns. Because $6\bQ_{K/2,n}=\mbf{G}$, where $\mbf{G}$ is a reciprocal lattice vector, this phase is only defined modulo $\pi/3$. In other words, the phases $\theta$ and $\theta+n\pi/3$ generate the same spatial pattern shifted by $n$ lattice sites along the direction parallel to $\bQ_{K/2,n}$. In \figref{fig: rhoK21 and QK21}, we present two characteristic choices for the phase of the condensate: $\theta=0$ and $\theta=\pi/6$.

In Tables~\ref{tab: transf. rho} and~\ref{tab: transf BDW}, we show the transformation properties of the CDW and BDW order parameters under point group symmetries, translations, and time reversal. An important observation is that the BDW order parameters are odd under reflections across an axis orthogonal to their wave vector, thereby enforcing a uniform onsite density. For $Q_{K/2,1}$, this operation corresponds to a reflection about the $x$-axis, given by $\sigma_v=C_{6z}^3\sigma_d$.

\subsubsection{Superconducting orders}\label{sec:U1_SC_orders}

\begin{table}[b]
    \centering
    \begin{tblr}{
        width = 0.45\linewidth,
        colspec = {c  X[c] X[c]},
        cells = {m},
        rowsep = {3pt}, 
        colsep = {0pt},
    }
    \toprule
    Symmetry & $\Delta_{d+id}$ & $\Delta_{d-id}$ \\\midrule

         $T_1$ &$\Delta_{d+id}$ & $\Delta_{d-id}$ \\

         $T_2$ &$\Delta_{d+id}$ & $\Delta_{d-id}$ \\

         $\sigma_d$ & $\Delta_{d-id}$ & $\Delta_{d+id}$ \\

         $C_{6z}$  & $e^{i\frac{2\pi}{3}}\Delta_{d+id}$ & $e^{-i\frac{2\pi}{3}}\Delta_{d-id}$ \\

         $\mathcal{T}$& $\Delta_{d-id}$ & $\Delta_{d+id}$ \\\bottomrule

    \end{tblr}
    \caption{Transformation properties of the uniform $\Delta_{d\pm id}$ superconducting order parameters.}
    \label{tab: transf d+id}
\end{table}

\begin{figure}[t!]
    \centering
    \includegraphics[width=0.75\textwidth]{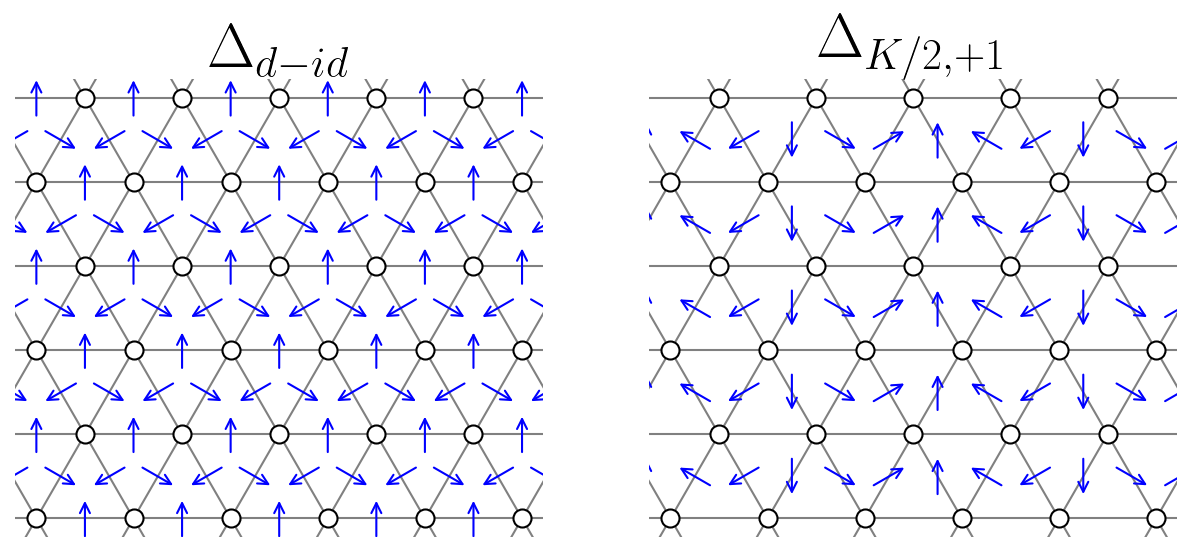}
    \caption{Spatial pattern realized by the condensation of $\Delta_{d-id}$ (left), and $\Delta_{K/2,+1}$ (right). The arrows represent the complex phase of the nearest neighbor pairing amplitudes. The absolute value of the pairing amplitude is uniform, except for bonds lacking an arrow, which indicates the absence of pairing on them.}
    \label{fig: Delta_K2_1}
\end{figure}

We now turn our attention to superconducting orders that break the global U(1) charge symmetry. We find two uniform chiral superconducting order parameters with the $d\pm i d$ symmetry:
\begin{subequations}
    \begin{align}
        &\Delta_{d+id} = e^{-i\frac{\pi}{3}}\left(B_{1,-}B_{2,+}+B_{3,+}B_{4,-}\right)\,,\\
        &\Delta_{d-id} = e^{+i\frac{\pi}{3}}\left(B_{1,+}B_{2,-}+B_{3,-}B_{4,+}\right)\,.
    \end{align}
\end{subequations}
Condensation of only of these two parameters leads to the additional spontaneous symmetry breaking of time-reversal symmetry and of mirror symmetry. Conversely, the simultaneous condensation of both with complex prefactors of equal absolute value induces nematic symmetry breaking without breaking time-reversal symmetry. In \tabref{tab: transf d+id}, we show the transformation properties of $\Delta_{d\pm id}$. In the left panel of \figref{fig: Delta_K2_1}, we show the phase pattern of $\Delta_{d-id}$.

We also find six superconducting order parameters with center-of-mass momenta $\pm\bQ_{K/2,n}$, defined by the bilinears:
\begin{subequations}
    \begin{align}
        &\Delta_{K/2,+1} = 2B_1^\top\kappa^1 B_4\,,\\
        &\Delta_{K/2,-1} = 2B_2^\top\kappa^1 B_3\,,\\
        &\Delta_{K/2,+2} = B_2^\top\kappa^1 B_2-B_3^\top\kappa^1 B_3\,,\\
        &\Delta_{K/2,-2} = B_1^\top\kappa^1 B_1-B_4^\top\,\kappa^1 B_4\,,\\
        &\Delta_{K/2,+3} = +i(B_1^\top\kappa^1 B_1+B_4^\top\,\kappa^1 B_4)\,,\\
        &\Delta_{K/2,-3} = -i(B_2^\top\kappa^1 B_2+B_3^\top\,\kappa^1 B_3)\,.
    \end{align}
\end{subequations}

\begin{table}[bh!]
    \centering
    \begin{tblr}{
        colspec = {c  X[c] X[c] X[c] X[c] X[c] X[c]},
        cells = {m},
        rowsep = {3pt}, 
        colsep = {0pt},
    }
    \toprule
    Symmetry & $\Delta_{K/2,+1}$ & $\Delta_{K/2,+2}$ & $\Delta_{K/2,+3}$ & $\Delta_{K/2,-1}$ & $\Delta_{K/2,-2}$ & $\Delta_{K/2,-3}$ \\\midrule

         $T_1$ & $e^{i\frac{2\pi}{3}} \Delta_{K/2,+1}$ & $e^{i\frac{\pi}{3}} \Delta_{K/2,+2}$ & $e^{-i\frac{\pi}{3}} \Delta_{K/2,+3}$ & $e^{-i\frac{2\pi}{3}} \Delta_{K/2,-1}$ & $e^{-i\frac{\pi}{3}} \Delta_{K/2,-2}$ & $e^{i\frac{\pi}{3}} \Delta_{K/2,-3}$\\

         $T_2$ & $e^{-i\frac{\pi}{3}} \Delta_{K/2,+1}$ & $e^{i\frac{\pi}{3}} \Delta_{K/2,+2}$ & $e^{i\frac{2\pi}{3}} \Delta_{K/2,+3}$ & $e^{i\frac{\pi}{3}} \Delta_{K/2,-1}$ & $e^{-i\frac{\pi}{3}} \Delta_{K/2,-2}$ & $e^{-i\frac{2\pi}{3}} \Delta_{K/2,-3}$\\

         $\sigma_d$ & $-\Delta_{K/2,-1}$ & $-\Delta_{K/2,+3}$ & $-\Delta_{K/2,+2}$ & $-\Delta_{K/2,+1}$ & $-\Delta_{K/2,-3}$ & $-\Delta_{K/2,-2}$\\

         $C_{6z}$  & $\Delta_{K/2,-3}$ & $\Delta_{K/2,+1}$ & $\Delta_{K/2,+2}$ & $\Delta_{K/2,+3}$ & $\Delta_{K/2,-1}$ & $\Delta_{K/2,-2}$ \\

         $\mathcal{T}$ & $\Delta_{K/2,-1}$ &$\Delta_{K/2,-2}$ &$\Delta_{K/2,-3}$ &$\Delta_{K/2,+1}$ &$\Delta_{K/2,+2}$ &$\Delta_{K/2,+3}$ \\\bottomrule

    \end{tblr}
    \caption{Transformation properties of the PDW order parameters at $\pm\bQ_{K/2,n}$.}
    \label{tab: transf. Delta K/2}
\end{table}
Condensation of a single bilinear gives rise to a pair density wave (PDW) phase of the Fulde--Ferrell (FF) type, where the pairing amplitude is spatially uniform but the phase spirals, as illustrated in the right panel of \figref{fig: Delta_K2_1}. In contrast, the simultaneous condensation of two bilinears at opposite wave vectors leads to a PDW of the Larkin--Ovchinnikov (LO) type. Here, the phase is uniform, but the amplitude is modulated with a spatial structure similar to that of $Q_{K/2,n}$ (shown in \figref{fig: rhoK21 and QK21}), inducing an additional BDW order at wave vectors $\vK$ and $\vK'$. The transformation properties of the order parameters $\Delta_{K/2,\pm n}$ are detailed in \tabref{tab: transf. Delta K/2}. A key property of these PDW orders is that they are odd under reflections across an axis orthogonal to the PDW wave vector.

\begin{figure}
    \centering
    \includegraphics[width=0.75\textwidth]{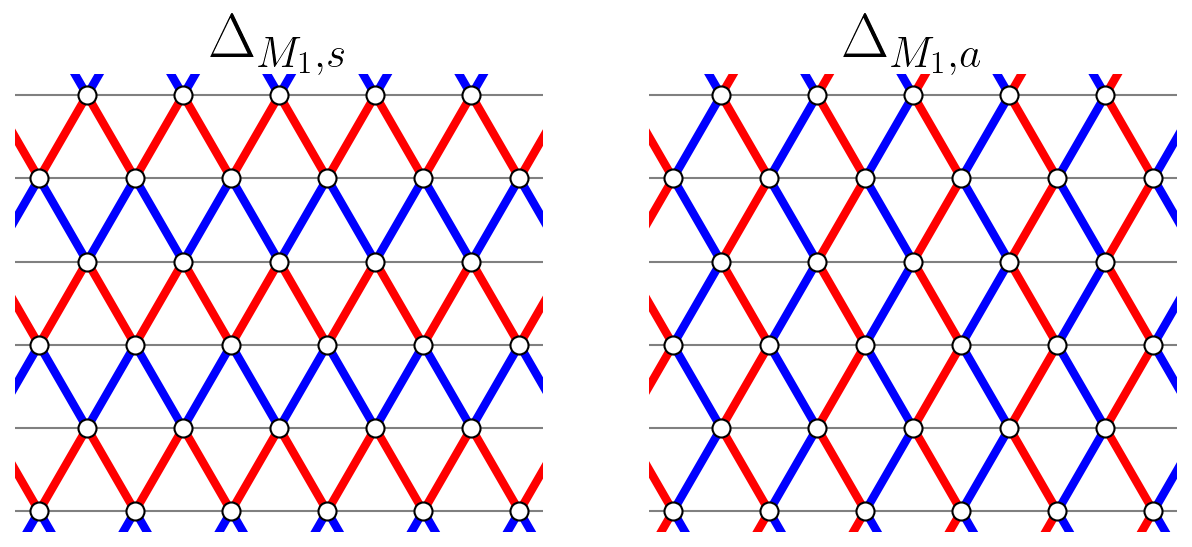}
    \caption{Sign structure of the pairing amplitude in the orders induced by the condensation of $\Delta_{M_1,s}$ (left) and $\Delta_{M_1,a}$ (right). Thick red bonds indicate a positive nearest-neighbor pairing amplitude, whereas thick blue bonds indicate a negative amplitude. Thin gray bonds indicate a vanishing pairing amplitude.}
    \label{fig: Delta M1}
\end{figure}

We also find two distinct types of PDWs characterized by the wave vectors $\bQ_{M_n}$:
\begin{subequations}
    \begin{align}
        &\Delta_{M_1,s} = B_1^\top e^{-i\frac{2\pi}{3}\kappa^3}\kappa^1 B_2-B_3^\top e^{i\frac{2\pi}{3}\kappa^3}\kappa^1 B_4\,,\\
        &\Delta_{M_2,s} = B_1^\top e^{-i\frac{\pi}{3}\kappa^3}\kappa^1 B_3+B_2^\top e^{i\frac{\pi}{3}\kappa^3}\kappa^1 B_4\,,\\
        &\Delta_{M_3,s} = -i\left(B_1^\top\kappa^1B_3-B_2^\top\kappa^1B_4\right)\,,\\
        & \nonumber \\
        &\Delta_{M_1,a} = B_1^\top e^{i\frac{5\pi}{6}\kappa^3}\kappa^1 B_2-B_3^\top e^{-i\frac{5\pi}{6}\kappa^3}\kappa^1 B_4\,,\\
        &\Delta_{M_2,a} = B_1^\top e^{-i\frac{5\pi}{6}\kappa^3}\kappa^1 B_3+B_2^\top e^{i\frac{5\pi}{6}\kappa^3}\kappa^1 B_4\,,\\
        &\Delta_{M_3,a} = -i\left(B_1^\top\kappa^2B_3+B_2^\top\kappa^2B_4\right)\,.
    \end{align}
\end{subequations}
Because $\bQ_{M_n}=-\bQ_{M_n}+\mbf{G}$, these PDW states exhibit only a modulation of the overall sign of the order parameter. Consequently, they never induce additional CDW order unless all three $\Delta_{M_n,s}$ or $\Delta_{M_n,a}$ condense simultaneously. The spatial modulations of the bond pairing amplitude, resulting from the condensation of $\Delta_{M_1,s}$ or $\Delta_{M_1,a}$, are illustrated in \figref{fig: Delta M1}. In \tabref{tab: transf. Delta M}, we show the transformation properties of these PDW order parameters. We note that each $\Delta_{M_n,s}$ order is symmetric under reflections across an axis parallel to its wave vector, whereas each $\Delta_{M_n,a}$ is antisymmetric under the same operation.

\begin{table}[b!]
    \centering
    \begin{tblr}{
        colspec = {c  X[c] X[c] X[c] X[c] X[c] X[c]},
        cells = {m},
        rowsep = {3pt}, 
        colsep = {0pt},
    }
    \toprule
    Symmetry & $\Delta_{M_1,s}$ & $\Delta_{M_2,s}$ & $\Delta_{M_3,s}$ & $\Delta_{M_1,a}$ & $\Delta_{M_2,a}$ & $\Delta_{M_3,a}$ \\ \midrule

         $T_1$ & $\Delta_{M_1,s}$ &  $-\Delta_{M_2,s}$ &  $-\Delta_{M_3,s}$ &  $\Delta_{M_1,a}$ &  $-\Delta_{M_2,a}$ &  $-\Delta_{M_3,a}$ \\

         $T_2$ & $-\Delta_{M_1,s}$ &  $-\Delta_{M_2,s}$ &  $\Delta_{M_3,s}$ &  $-\Delta_{M_1,a}$ &  $-\Delta_{M_2,a}$ &  $\Delta_{M_3,a}$ \\

         $\sigma_d$ & $\Delta_{M_1,s}$ &  $\Delta_{M_3,s}$ &  $\Delta_{M_2,s}$ &  $-\Delta_{M_1,a}$ &  $-\Delta_{M_3,a}$ &  $-\Delta_{M_2,a}$ \\

         $C_{6z}$  &  $-\Delta_{M_3,s}$ &  $-\Delta_{M_1,s}$ &  $-\Delta_{M_2,s}$ &  $-\Delta_{M_3,a}$ &  $-\Delta_{M_1,a}$ &  $-\Delta_{M_2,a}$ \\

         $\mathcal{T}$ &  $\Delta_{M_1,s}$ & $\Delta_{M_2,s}$ & $\Delta_{M_3,s}$ & $\Delta_{M_1,a}$ & $\Delta_{M_2,a}$ & $\Delta_{M_3,a}$ \\\bottomrule

    \end{tblr}
    \caption{Transformation properties of the PDW order parameters at $\bQ_{M_n}$.}
    \label{tab: transf. Delta M}
\end{table}

\subsubsection{Current orders}
\begin{figure}[b]
    \centering
    \includegraphics[width=0.375\textwidth]{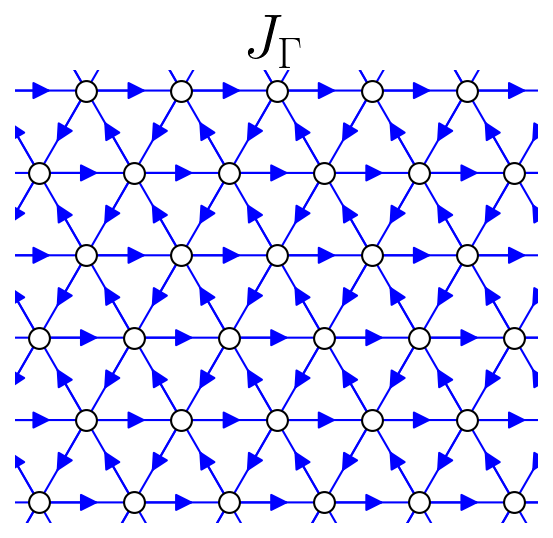}
    \caption{Current patterns realized by the condensation of $J_\Gamma$. The arrows on each bond represent the direction in which the current flows.}
    \label{fig: J gamma}
\end{figure}

We now turn our attention to current orders. We find a uniform current order, described by the bilinear
\begin{equation}
    J_\Gamma = |B_1|^2-|B_2|^2-|B_3|^2+|B_4|^2\,,
\end{equation}
which creates opposite fluxes in the two inequivalent triangular plaquettes of the triangular lattice. The spatial current patterns are shown in \figref{fig: J gamma}.

\begin{figure}
    \centering
    \includegraphics[width=0.75\textwidth]{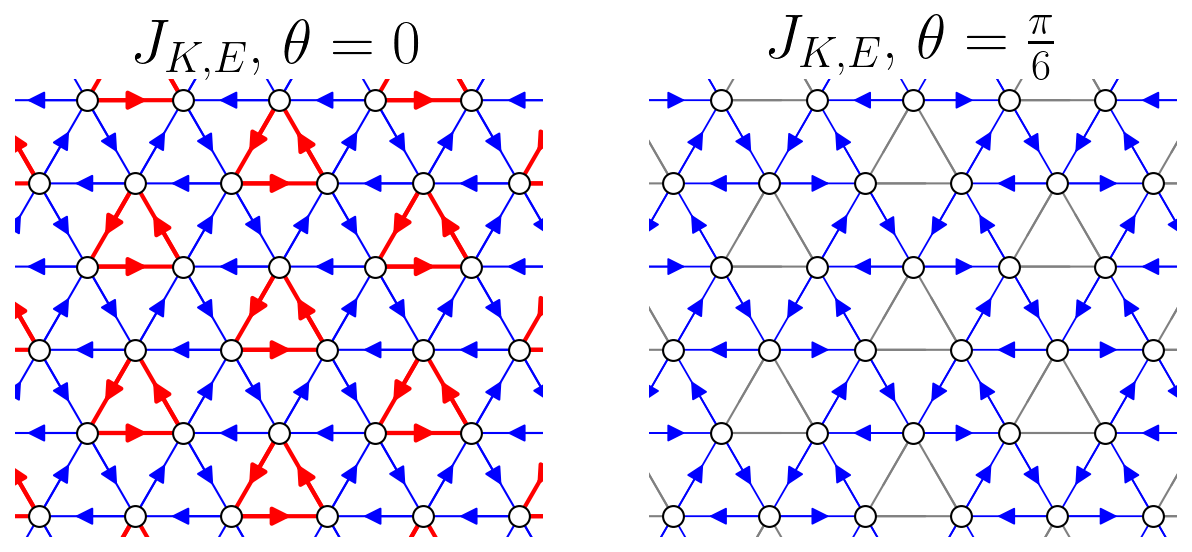}
    \caption{Current patterns realized by the condensation of $J_{K,E}$ for a complex phase of $0$ (left) and $\pi/6$ (right). The arrows on each bond represent the direction of current flow. Red bonds indicate currents twice as intense as those on blue bonds, and thin gray bonds are bonds on which no current flows. Spatial patterns for $J_{K,E^*}$ can be obtained from those above by reflecting across the horizontal axis.}
    \label{fig: JKE}
\end{figure}

We then find two symmetry-related current loop orders at $\bQ_K=(4\pi/3,0)$:
\begin{subequations}
    \begin{align}
        &J_{K,E^{\phantom{*}}} = e^{i\frac{\pi}{6}}\left(B^*_{1,+}B_{3,+}-B^*_{4,+}B_{2,+}\right)\,,\\
        &J_{K,E^*} = -e^{-i\frac{\pi}{6}}\left(B^*_{1,-}B_{3,-}-B^*_{4,-}B_{2,-}\right)\,.
    \end{align}
\end{subequations}
\begin{table}[b!]
    \centering
    \begin{tblr}{
        width = 0.6\linewidth,
        colspec = {c  X[c] X[c] X[c]},
        cells = {m},
        rowsep = {3pt}, 
        colsep = {0pt},
    }
    \toprule
    Symmetry & $J_\Gamma$ & $J_{K,E}$ & $J_{K,E^*}$\\ \midrule
         $T_1$ & $J_\Gamma$ & $e^{-i\frac{2\pi}{3}} J_{K,E}$ & $e^{-i\frac{2\pi}{3}} J_{K,E^*}$\\

         $T_2$ & $J_\Gamma$ &  $e^{-i\frac{2\pi}{3}} J_{K,E}$ & $e^{-i\frac{2\pi}{3}} J_{K,E^*}$\\

         $\sigma_d$ & $-J_\Gamma$ & $-J_{K,E}^*$ & $-J_{K,E^*}^*$ \\

         $C_{6z}$  & $-J_\Gamma$ & $e^{i\frac{\pi}{3}}J_{K,E^*}^*$ & $e^{-i\frac{\pi}{3}}J_{K,E}^*$\\

         $\mathcal{T}$ & $-J_\Gamma$ & $-J_{K,E}^*$ & $-J_{K,E^*}^*$  \\\bottomrule

    \end{tblr}
    \caption{Transformation properties of the uniform current order parameters and of those at $\bQ_K$.}
    \label{tab: transf. J Gamma K}
\end{table}

Note that both $J_{K,E}$ and $J_{K,E^*}$ are complex order parameters, and their complex phases influence the spatial current patterns induced by these orders. In \figref{fig: JKE} we show the spatial patterns of $J_{K,E}$ for a complex phase of $0$ and $\pi/6$. In \tabref{tab: transf. J Gamma K}, we show the transformation properties of $J_\Gamma$, $J_{K,E}$ and $J_{K,E^*}$.

\begin{figure}
    \centering
    \includegraphics[width=0.75\textwidth]{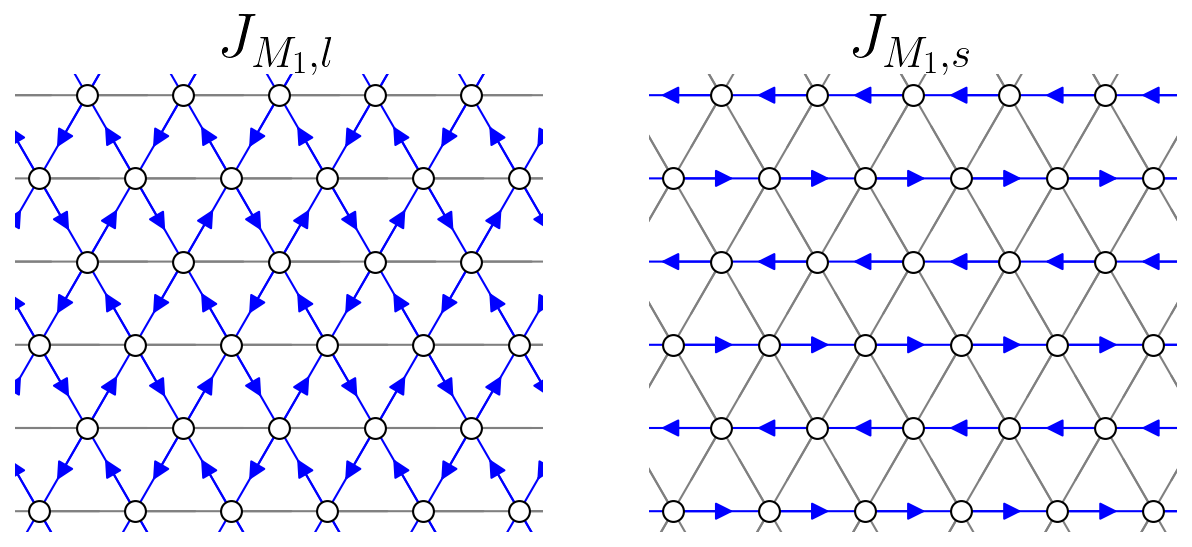}
    \caption{Current patterns realized by the condensation of $J_{M_1,l}$ (left) and $J_{M_1,s}$ (right). The arrows on each bond represent the direction in which the current flows, and thin gray lines indicate bonds on which no current flows.}
    \label{fig: JM1}
\end{figure}

Finally, we find two types of current loop orders at $\bQ_{M_n}$, defined by the bilinears:
\begin{subequations}
    \begin{align}
        &J_{M_1,l} = |B_1|^2-|B_2|^2+|B_3|^2-|B_4|^2\,,\\
        &J_{M_2,l} = -B_1^\dagger B_4-B^\dagger_4 B_1+B^\dagger_2 B_3+B^\dagger_3 B_2\,,\\
        &J_{M_3,l} = i\left(B_1^\dagger B_4-B^\dagger_4 B_1+B^\dagger_2 B_3-B^\dagger_3 B_2\right)\,,\\
        &\nonumber\\
        &J_{M_1,s} = B_1^\dagger\,\kappa^3B_1-B_2^\dagger\,\kappa^3B_2+B_3^\dagger\,\kappa^3B_3-B_4^\dagger\,\kappa^3B_4\,,\\
        &J_{M_2,s} = -B_1^\dagger\,\kappa^3 B_4-B^\dagger_4 \,\kappa^3B_1+B^\dagger_2\,\kappa^3 B_3+B^\dagger_3 \,\kappa^3B_2\,,\\
        &J_{M_3,s} = i\left(B_1^\dagger \,\kappa^3B_4-B^\dagger_4 \,\kappa^3B_1+B^\dagger_2\,\kappa^3 B_3-B^\dagger_3\,\kappa^3 B_2\right)\,.
    \end{align}
\end{subequations}

The current patterns induced by the condensation of $J_{M_1,l}$ and $J_{M_1,s}$ are shown in \figref{fig: JM1}. We see that while $J_{M_n,l}$ creates loop currents around square plaquettes (consisting of two triangular plaquettes), $J_{M_n,s}$ induces staggered currents. \tabref{tab: transf. J M} shows the transformation properties of these current orders. The main difference between $J_{M_1,l}$ and $J_{M_1,s}$ is that the former are odd under reflections across an axis orthogonal to $\bQ_{M_n}$, while the latter are even. 

\begin{table}[]
    \centering
    \begin{tblr}{
        colspec = {c  X[c] X[c] X[c] X[c] X[c] X[c]},
        cells = {m},
        rowsep = {3pt}, 
        colsep = {0pt},
    }
    \toprule
    Symmetry & $J_{M_1,l}$ &$J_{M_2,l}$ &$J_{M_3,l}$ &$J_{M_1,s}$ &$J_{M_2,s}$ &$J_{M_3,s}$ \\ \midrule

         $T_1$ & $J_{M_1,l}$ &  $-J_{M_2,l}$ &  $-J_{M_3,l}$ &  $J_{M_1,s}$ &  $-J_{M_2,s}$ &  $-J_{M_3,s}$ \\

         $T_2$ & $-J_{M_1,l}$ &  $-J_{M_2,l}$ &  $J_{M_3,l}$ &  $-J_{M_1,s}$ &  $-J_{M_2,s}$ &  $J_{M_3,s}$\\

         $\sigma_d$ & $-J_{M_1,l}$& $-J_{M_3,l}$& $-J_{M_2,l}$&$-J_{M_1,s}$& $-J_{M_3,s}$& $-J_{M_2,s}$\\
         
         $C_{6z}$ & $J_{M_3,l}$& $J_{M_1,l}$& $J_{M_2,l}$& $-J_{M_3,s}$& $-J_{M_1,s}$& $-J_{M_2,s}$ \\
         
         $\mathcal{T}$ & $-J_{M_1,l}$ &$-J_{M_2,l}$ &$-J_{M_3,l}$ &$-J_{M_1,s}$ &$-J_{M_2,s}$ &$-J_{M_3,s}$ \\\bottomrule
    \end{tblr}
    \caption{Transformation properties of the current order parameters at $\bQ_{M_n}$.}
    \label{tab: transf. J M}
\end{table}

\subsubsection{Field theory}

We now derive a continuum field theory for the fractionalized order parameters within the Dirac spin liquid:
\begin{equation}
    \mathcal{L}=\sum_{\eta=1}^4|D_\mu B_\eta|^2 + r\rho_\Gamma+V(B_\eta)\,,
\end{equation}
where the covariant derivative is defined as $D_\mu=\partial_\mu + i \mathcal{A}_\mu \tau^3$, with the internal U(1) gauge field $\mathcal{A}_\mu$, and $\rho_\Gamma=\sum_\eta |B_\eta|^2$. The potential $V(B_\eta)$ can be constructed by combining all bilinears discussed in the previous sections. At the quartic level, this leads to:
\begin{equation}
    \begin{aligned}
        V(B_\eta) = &u\rho_\Gamma^2 + v_1\sum_{n=1}^3 \rho_{M_n}^2 + v_2\sum_{n=1}^3 |\rho_{K/2,n}|^2 + v_3\sum_{n=1}^3 Q_{M_n}^2+ v_4\sum_{n=1}^3 |Q_{K/2,n}|^2\\
        &+v_5\left[|\Delta_{d+id}|^2+|\Delta_{d-id}|^2\right]+v_6\sum_{n\in\{\pm1,\pm2,\pm3\}}|\Delta_{K/2,n}|^2 + v_7\sum_{n=1}^3 |\Delta_{M_n,s}|^2+ v_8\sum_{n=1}^3 |\Delta_{M_n,a}|^2\\
        &+v_9 J_\Gamma^2 + v_{10}\left[|J_{K,E}|^2+|J_{K,E^*}|^2\right] + v_{11}\sum_{n=1}^3 J_{M_n,l}^2 + v_{12}\sum_{n=1}^3 J_{M_n,s}^2\,.
    \end{aligned}
\end{equation}
Note that $u$ and $v_{1,\dots,12}$ are not all linearly independent. Because these bilinears are constructed from the eight-component complex field $B_{\eta,\alpha}$, they form the gauge-invariant subspace of $8\times 8$ matrices. Consequently, any product of two bilinears can be mapped onto a linear combination of other quartic terms via the completeness relations of the matrix basis. Applying these Fierz identities~\cite{Fierz_1937} algebraically removes the redundant terms. 

Alternatively, one can obtain the identical field theory by constructing all gauge- and charge-invariant quadratic and quartic polynomials in $B_{\eta,\alpha}$ that are invariant under the lattice and time reversal symmetry.

Both methods lead to a field theory with only $8$ independent parameters out of the $13$:
\begin{equation}\label{eq:Fierz corrected potential U(1)DSL}
    \begin{aligned}
        V(B_\eta) = &\tilde{u}\rho_\Gamma^2 + \tilde{v}_1\sum_{n=1}^3 \rho_{M_n}^2 + \tilde{v}_2\sum_{n=1}^3 Q_{M_n}^2+\tilde{v}_3\sum_{n\in\{\pm1,\pm2,\pm3\}}|\Delta_{K/2,n}|^2 + \tilde{v}_4|J_{\Gamma}|^2\\ &+ \tilde{v}_5[|\Delta_{d+id}|^2+|\Delta_{d-id}|^2]+ \tilde{v}_{6}\sum_{n=1}^3 J_{M_n,l}^2 + \tilde{v}_7\sum_{n=1}^3 J_{M_n,s}^2\,.
    \end{aligned}
\end{equation}

\subsubsection{Representative condensates}
Here, we review several possible phases that can emerge from the theory presented in \eqnref{eq:Fierz corrected potential U(1)DSL}.

\begin{figure}[t!]
    \centering
    \includegraphics[width=0.75\linewidth]{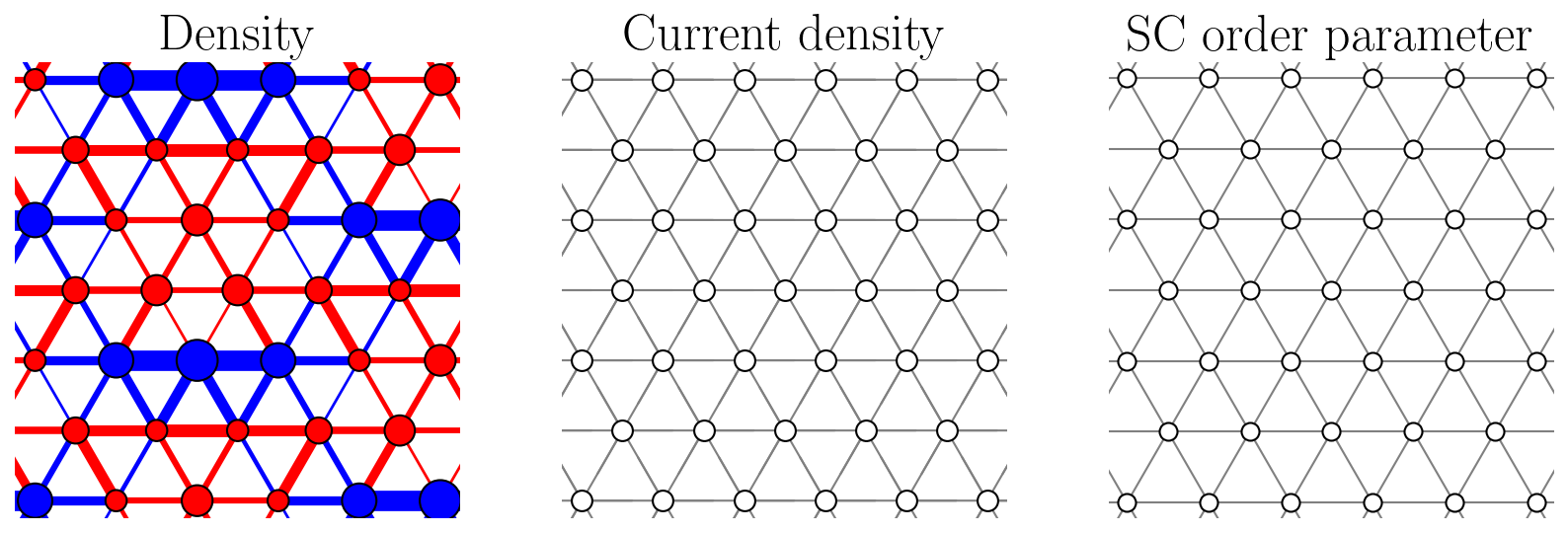}
    \caption{(a) Charge and bond, (b) current, and (c) pairing density profiles for Phase A. The conventions are the same as in the previous figures.}
    \label{fig:phaseA}
\end{figure}

\begin{figure}[t!]
    \centering
    \includegraphics[width=0.75\linewidth]{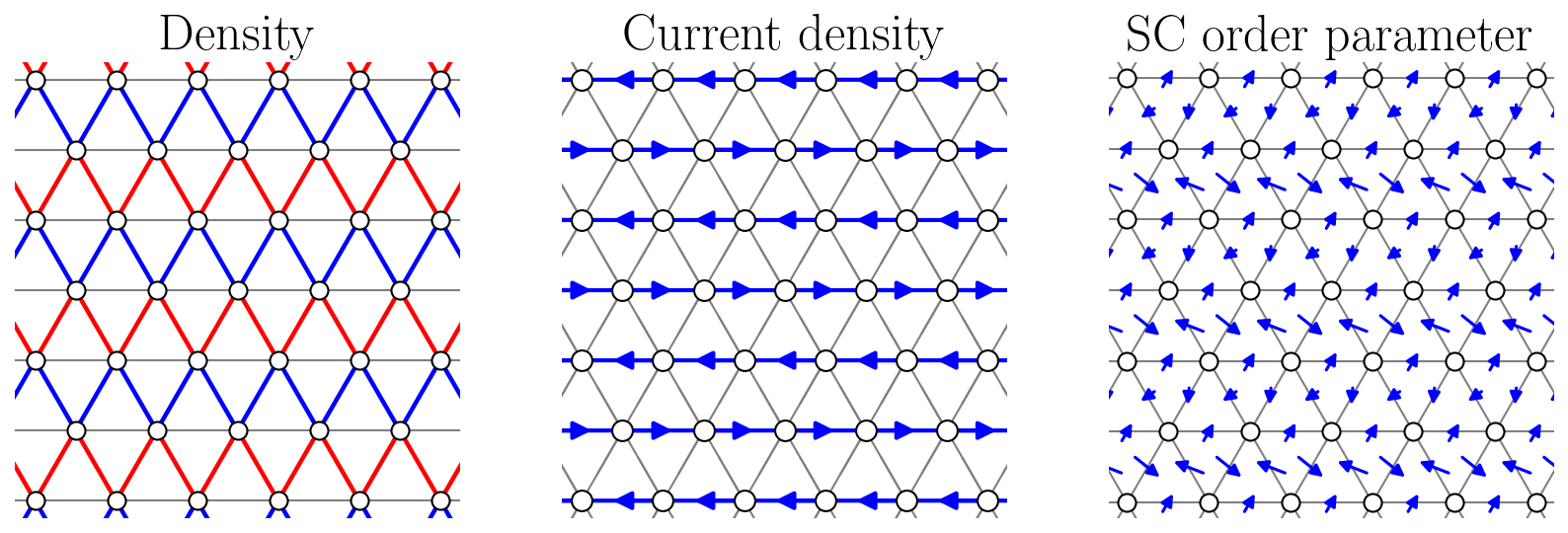}
    \caption{(a) Charge and bond, (b) current, and (c) pairing density profiles for Phase B. The conventions are the same as in the previous figures.}
    \label{fig:phaseB}
\end{figure}

\begin{figure}[t!]
    \centering
    \includegraphics[width=0.75\linewidth]{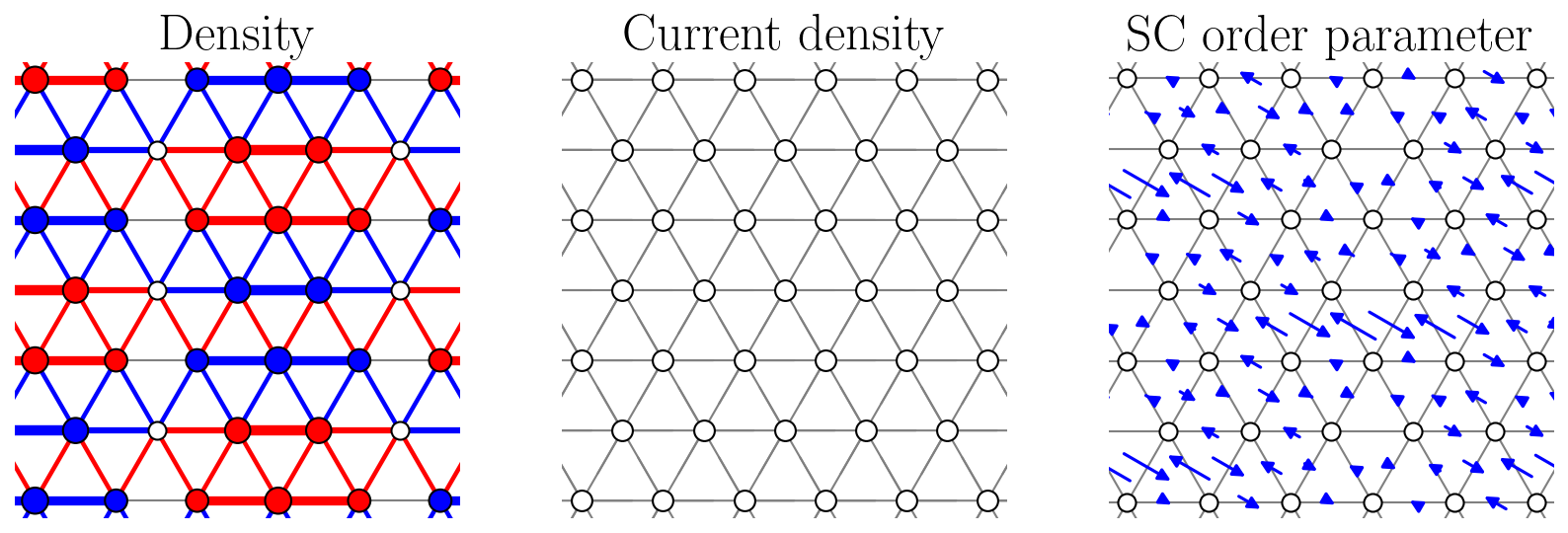}
    \caption{(a) Charge and bond, (b) current, and (c) pairing density profiles for Phase C. The conventions are the same as in the previous figures.}
    \label{fig:phaseC}
\end{figure}

\begin{figure}[t!]
    \centering
    \includegraphics[width=0.75\linewidth]{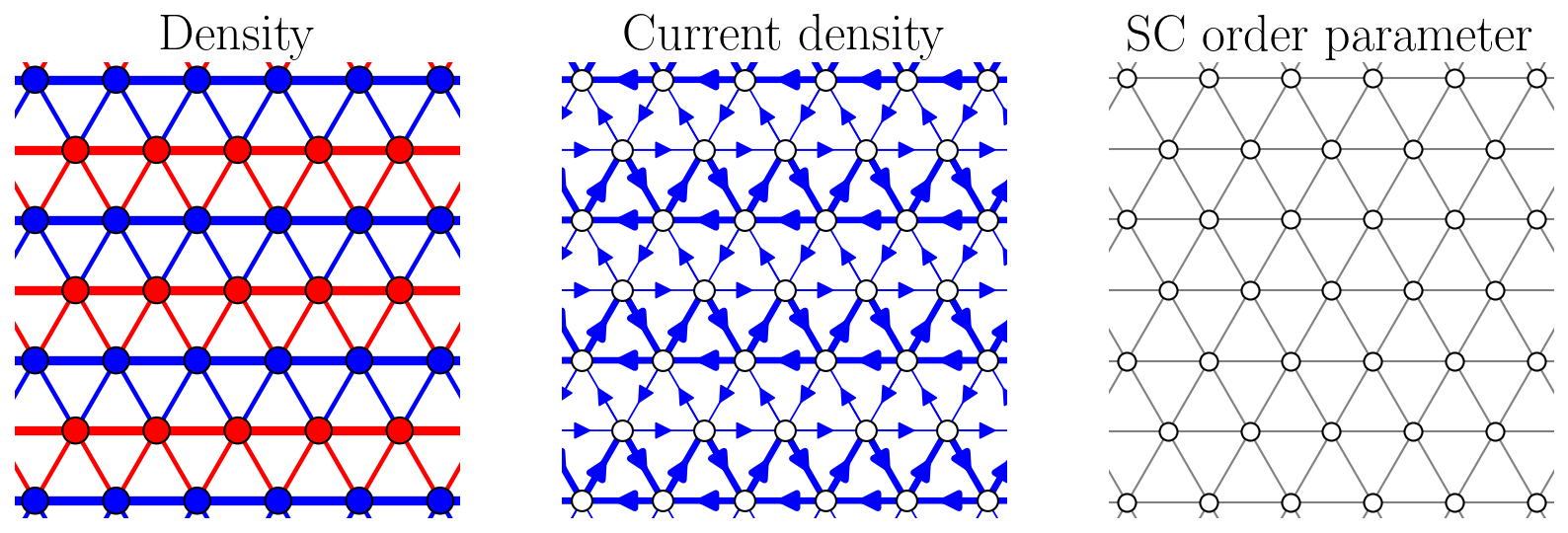}
    \caption{(a) Charge and bond, (b) current, and (c) pairing density profiles for Phase D. The conventions are the same as in the previous figures.}
    \label{fig:phaseD}
\end{figure}

\begin{figure}[t!]
    \centering
    \includegraphics[width=0.75\linewidth]{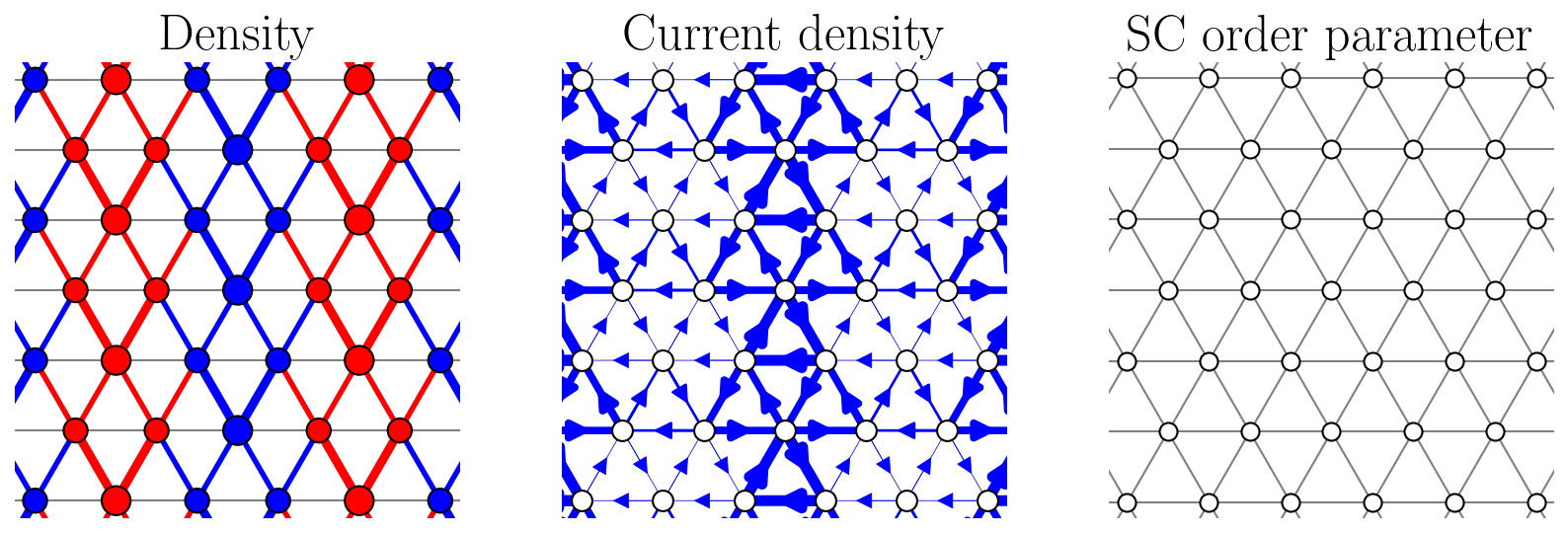}
    \caption{(a) Charge and bond, (b) current, and (c) pairing density profiles for Phase E. The conventions are the same as in the previous figures.}
    \label{fig:phaseE}
\end{figure}

\begin{itemize}
    \item \textbf{Phase A: Translation symmetry-breaking state.} This state possesses a $\sqrt{12}\times\sqrt{12}$ supercell. The fields are given by $B_1=\frac{b}{\sqrt{2}}(1,0)$, $B_2=\frac{b}{\sqrt{2}}(e^{i\frac{\pi}{12}},0)$, and $B_3=B_4=0$. The charge and bond density profiles are shown in \figref{fig:phaseA}.

    \item \textbf{Phase B: Translation symmetry-breaking chiral superconducting state.} This state possesses a $1\times 2$ supercell. The fields are given by $B_1=\frac{b}{\sqrt{2}}(1,0)$, $B_2=\frac{b}{\sqrt{2}}(0,1)$, and $B_3=B_4=0$. The charge, bond, current, and pairing density profiles are shown in \figref{fig:phaseB}.

    \item \textbf{Phase C: Translation symmetry-breaking nematic superconducting state.} This state possesses a $\sqrt{12}\times\sqrt{12}$ supercell. The fields are given by $B_1=\frac{b}{2}(1,e^{-i\frac{5\pi}{12}})$, $B_2=\frac{b}{2}(e^{i\frac{\pi}{12}},1)$, and $B_3=B_4=0$. The charge, bond, current, and pairing density profiles are shown in \figref{fig:phaseC}.

    \item \textbf{Phase D: Current loop state with charge and bond density modulations.} This state possesses a $1\times 2$ supercell. The fields are given by $B_1=b(1,0)$ and $B_2=B_3=B_4=0$. The charge, bond, current, and pairing density profiles are shown in \figref{fig:phaseD}.

    \item \textbf{Phase E: Current loop state with charge and bond density modulations.} This state possesses a $3\times 2$ supercell. The fields are given by $B_1=\frac{b}{\sqrt{2}}(1,0)$, $B_3=\frac{b}{\sqrt{2}}(e^{-i\frac{\pi}{6}},0)$, and $B_2=B_4=0$. The charge, bond, current, and pairing density profiles are shown in \figref{fig:phaseE}.
\end{itemize}

We note that we could not find uniform superconducting states by minimizing all other orders over the space of condensation patterns $\langle B_{\eta,\alpha}\rangle$ with finite $\Delta_{d\pm id}$.

\subsubsection{A minimal lattice realization}

\begin{figure}[t!]
    \centering
    \includegraphics[width=0.95\linewidth]{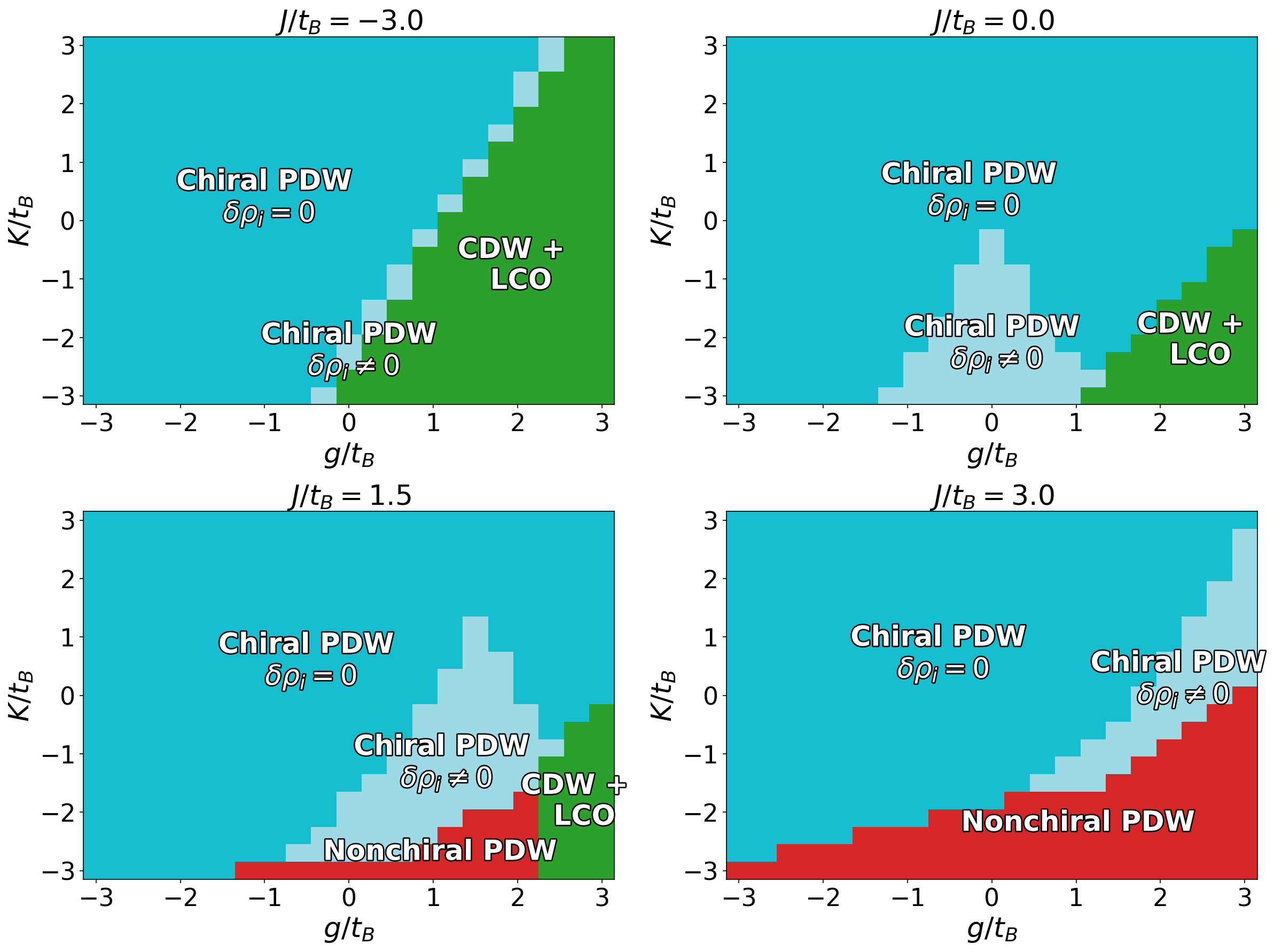}
    \caption{Phase diagrams obtained by minimizing the lattice functional
    in \eqnref{eq:lattice-functional} within the low-energy chargon
    subspace of \eqnref{eq: continuum expansion Bia}. We fix
    $r=2t_B$, $u=10t_B$, and $t_B>0$, and show representative cuts in $g$ and $K$ at the
    indicated values of $J/t_B$. The phases are classified by their
    gauge-invariant charge, current, and superconducting bilinears.
    Here $\delta\rho_i=\rho_i-\overline{\rho}$, PDW denotes a pair-density wave, and LCO denotes loop-current order.}
    \label{fig:lattice-phases}
\end{figure}
To illustrate which of the symmetry-allowed chargon condensates can be
stabilized by a simple short-range model, we complement the continuum
analysis with a variational minimization of the static lattice functional
\begin{align}
\mathcal{F}_{\mathrm{lat}}
={}&
t_B
\sum_{\langle i,j\rangle}
\left(
B_i^\dagger e_{ij}\tau^3 B_j+\mathrm{H.c.}
\right)
+r\sum_i B_i^\dagger B_i
+u\sum_i\left(B_i^\dagger B_i\right)^2
\nonumber\\
&+
K\sum_{\langle i,j\rangle}Q_{ij}^2
+J\sum_{\langle i,j\rangle}J_{ij}^2
+g\sum_{\langle i,j\rangle}|\Delta_{ij}|^2 .
\label{eq:lattice-functional}
\end{align}

Here, $e_{ij}=\pm1$ denotes the nearest-neighbor sign structure of the
$0$--$\pi$ flux state, and $Q_{ij}$, $J_{ij}$, and $\Delta_{ij}$ are the
gauge-invariant bond, current, and pairing bilinears defined in \eqnref{eq: gauge inv bilinears}. The local quartic interaction with $u>0$
stabilizes the amplitude of the chargon condensate. Negative values of
$K$, $J$, or $g$ favor, respectively, bond-density, current, or
superconducting correlations. These tendencies should not, however, be
regarded as independent: all three observables are bilinears of the same
chargon field, and a condensate favored in one channel generally induces
additional orders in the other channels.

We fix $t_B>0$, $r=2t_B$, and $u=10t_B$, and restrict the chargon field
to the low-energy subspace of \eqnref{eq: continuum expansion Bia}. The eight
complex amplitudes $B_{\eta,a}$ are then treated as variational
parameters. This projection retains all condensates constructed from the
four chargon minima, but excludes higher-energy chargon modes and spatial
textures that cannot be represented within the low-energy expansion.

The resulting phase diagrams are shown in \figref{fig:lattice-phases}
for several representative values of $J/t_B$. We classify the minima
using their gauge-invariant observables. Four principal types of states
are obtained. The first is a chiral pair-density wave with a uniform
onsite density, $\delta\rho_i\equiv\rho_i-\overline{\rho}=0$. Although
the onsite density is uniform, this state can possess modulated bond
observables and breaks time-reversal symmetry. The second is a chiral
pair-density wave accompanied by a nonzero charge-density modulation,
$\delta\rho_i\neq0$. The third is a nonchiral pair-density wave, which
preserves time reversal but breaks translations and, in general, lattice
rotation symmetry. Finally, we find nonsuperconducting states in which a
charge-density wave coexists with loop-current order, denoted
CDW+LCO in \figref{fig:lattice-phases}.

The overall evolution of the phase diagram is consistent with the roles
of the three interaction parameters. A more attractive pairing
interaction, corresponding to decreasing $g$, generally stabilizes the
PDW phases. Decreasing $K$ enhances density and bond modulations and can
convert a PDW with uniform onsite density into one with
$\delta\rho_i\neq0$, or, when pairing is sufficiently disfavored, into
the CDW+LCO state. Increasing $J$ penalizes current order: the CDW+LCO
region consequently shrinks, while a nonchiral PDW appears over an
increasing part of the phase diagram. Nevertheless, the phase boundaries
cannot be inferred from the signs of $K$, $J$, and $g$ alone, because the
different physical orders are constrained composites of the same
fractionalized field. In particular, superconducting order may remain
nonzero even for moderately positive $g$.

Representative real-space configurations of condensates closely related
to the phases found in \figref{fig:lattice-phases} are shown in
Figs.~\ref{fig:phaseB}--\ref{fig:phaseE}. The configurations shown here are not, in general, the exact minima found
in \figref{fig:lattice-phases}, but they provide simple representatives
of phases with the same dominant orders and symmetry-breaking patterns. In this sense, Phase B illustrates a chiral
PDW with uniform onsite density, Phase C a nonchiral nematic PDW, and
Phases D and E two CDW+LCO states with distinct translation-breaking
patterns. Phase A provides a further example of a symmetry-allowed
density-wave condensate, although no closely corresponding phase is
selected in the parameter cuts shown in \figref{fig:lattice-phases}.

An important outcome of this minimal calculation is that none of the
displayed parameter regimes stabilizes a spatially uniform
superconductor. Although uniform $d\pm id$ pairing is allowed by the
symmetry analysis, the short-range functional considered here instead
selects finite-momentum superconductivity, frequently intertwined with
density or current order. This absence is not a general prohibition:
longer-range interactions, additional quartic invariants, fluctuations,
or chargon modes away from the band minima may stabilize the uniform
superconducting states. 

\subsection{Symmetry-breaking patterns of the \texorpdfstring{$\Z_2$}{Z2} spin liquid}
\label{sec:Z2}

In the $\Z_2$ spin liquid, the mixing angle $\mix$ is non-zero, which distinguishes it from the $\U(1)$ Dirac spin liquid. This finite $\mix$ alters the explicit form of the bilinears and allows for new CDW and superconducting orders that were strictly prohibited by the chargon mode structure in the $\U(1)$ DSL. It is crucial to emphasize that all other fractionalized orders---including the bond density waves, current loop orders, and pair density waves detailed previously---are still present in the $\Z_2$ phase with altered bilinear representations. However, because we are interested in the primary experimental signatures of this phase transition, we focus on the most phenomenologically relevant order parameters in the following: the CDWs and the uniform SC orders.

\subsubsection{CDW orders}

The non-zero $\mix$ permits the formation of $K$-point charge orders, which vanish in the $\U(1)$ spin liquid. The $K$-points are given by $\vK=-\vK'=(4\pi/3,0)$:
\begin{subequations}
\begin{align}
    \rho_{K}&=\sin(\mix)\left(-B_{2}^\dagger\kappa^2B_{4}+B_3^\dagger\kappa^2B_1\right)\,,\\
    \rho_{K'}&=\sin(\mix)\left(-B_{4}^\dagger\kappa^2B_{2}+B_1^\dagger\kappa^2B_3\right)\,.
\end{align}
\end{subequations}

For the $M$- and $K/2$-point CDWs, the low-energy representations are modified to:
\begin{subequations}
    \begin{align}
        &\rho_{M_1}=\sum_{\eta=1}^4B^\dagger_\eta [(\sin(\mix)/\sqrt{2}\kappa^1+\kappa^3)(-1)^{\lfloor \frac{\eta-1}{2}\rfloor}+(-1)^{\eta-1}\sin(\mix)\sqrt{3/2}\kappa^2] B_{\eta}\\
        &\rho_{M_2}=-\Bigg[B^\dagger_{1}\left(\frac{\sin(\mix)}{\sqrt{2}}\kappa^1-\frac{\sin(\mix)\sqrt{3}}{\sqrt{2}}\kappa^2+\kappa^3\right)B_{4}+B^\dagger_{4}\left(\frac{\sin(\mix)}{\sqrt{2}}\kappa^1-\frac{\sin(\mix)\sqrt{3}}{\sqrt{2}}\kappa^2+\kappa^3\right)B_{1}\nonumber\\
        &\qquad\quad+B^\dagger_{3}\left(\frac{\sin(\mix)}{\sqrt{2}}\kappa^1+\frac{\sin(\mix)\sqrt{3}}{\sqrt{2}}\kappa^2+\kappa^3\right)B_{2}+B^\dagger_{2}\left(\frac{\sin(\mix)}{\sqrt{2}}\kappa^1+\frac{\sin(\mix)\sqrt{3}}{\sqrt{2}}\kappa^2+\kappa^3\right)B_{3}\Bigg]\,,\\
        & \rho_{M_3}=iB^\dagger_{1}(\kappa^3-\sqrt{2}\sin(\mix)\kappa^1)B_{4}+iB^\dagger_{3}(\kappa^3-\sqrt{2}\sin(\mix)\kappa^1)B_{2}+\mathrm{H.c.}\\
        \nonumber\\
        &\rho_{K/2,1}=B_3^\dagger \left(e^{-i\frac{\pi}{6}\kappa^3}+i\frac{\sin(\mix)}{\sqrt{2}}\kappa^1\right) B_1+B_2^\dagger \left(e^{-i\frac{\pi}{6}\kappa^3}+i\frac{\sin(\mix)}{\sqrt{2}}\kappa^1\right)B_4\,,\\
        &\rho_{K/2,2}=B_1^\dagger\left(e^{i\frac{5\pi}{6}\kappa^3}-i\frac{\sin(\mix)}{\sqrt{2}}\kappa^1\right)B_2
        -B_4^\dagger\left(e^{i\frac{5\pi}{6}\kappa^3}-i\frac{\sin(\mix)}{\sqrt{2}}\kappa^1\right)B_3\,,\\
        &\rho_{K/2,3}=-B^\dagger_2\left(\kappa^3+\frac{\sqrt{3}\sin(\mix)}{\sqrt{2}}\kappa^1\right) B_1 - B^\dagger_3 \left(\kappa^3+\frac{\sqrt{3}\sin(\mix)}{\sqrt{2}}\kappa^1\right) B_4\,.
    \end{align}
\end{subequations}

The symmetry transformations of these order parameters are shown in \tabref{tab: transf. rho Z2}.

\begin{table}
    \centering
    \begin{tblr}{
        colspec = {c  X[c] X[c] X[c] X[c] X[c] X[c] X[c] X[c]},
        cells = {m},
        rowsep = {3pt}, 
        colsep = {0pt},
    }
    \toprule
    Symmetry & $\rho_K$ & $\rho_{K'}$ & $\rho_{M_1}$ & $\rho_{M_2}$ & $\rho_{M_3}$ 
        & $\rho_{K/2,1}$ & $\rho_{K/2,2}$ & $\rho_{K/2,3}$ \\\midrule
         $T_1$ & $e^{i\frac{2\pi}{3}}\rho_K $ & $ e^{-i\frac{2\pi}{3}}\rho_{K'}$ & $\rho_{M_1}$ & $-\rho_{M_2}$ & $-\rho_{M_3}$ 
         & $e^{i\frac{2\pi}{3}}\rho_{K/2,1}$ & $e^{i\frac{\pi}{3}}\rho_{K/2,2}$ &  $e^{-i\frac{\pi}{3}}\rho_{K/2,3}$ \\
         $T_2$ & $e^{i\frac{2\pi}{3}}\rho_K $ & $ e^{-i\frac{2\pi}{3}}\rho_{K'}$ & $-\rho_{M_1}$ &  $-\rho_{M_2}$ & $\rho_{M_3}$ 
         &  $e^{-i\frac{\pi}{3}}\rho_{K/2,1}$ & $e^{i\frac{\pi}{3}}\rho_{K/2,2}$ & $e^{i\frac{2\pi}{3}}\rho_{K/2,3}$\\
         $\sigma_d$ & $\rho_{K'} $ & $\rho_{K} $ & $\rho_{M_1}$ & $\rho_{M_3}$ & $\rho_{M_2}$ 
         & $\rho_{K/2,1}^*$ & $\rho_{K/2,3}$ & $\rho_{K/2,2}$\\
         $C_{6z}$  & $ \rho_{K'} $ & $ \rho_K$ & $\rho_{M_3}$ &$\rho_{M_1}$& $\rho_{M_2}$
         & $\rho_{K/2,3}^*$ & $\rho_{K/2,1}$ & $\rho_{K/2,2}$\\
         $\mathcal{T}$ & $ \rho_{K'}$ & $\rho_{K} $ & $\rho_{M_1}$ &$\rho_{M_2}$ & $\rho_{M_3}$ 
         & $\rho_{K/2,1}^*$ & $\rho_{K/2,2}^*$ & $\rho_{K/2,3}^*$ \\\bottomrule
    \end{tblr}
    \caption{Transformation properties of the CDW order parameters emerging from the $\Z_2$ spin liquid. While the $\rho_K$ and $\rho_{K'}$ orders are newly allowed by the non-zero mixing angle $\mix$, the projective transformations of the $M$- and $K/2$-point CDWs remain identical to the $\U(1)$ spin liquid case.}
    \label{tab: transf. rho Z2}
\end{table}

\subsubsection{Superconducting orders}

We now examine the uniform superconducting orders that break the global $\U(1)$ charge symmetry. We find two chiral superconducting order parameters with $d\pm i d$ symmetry, analogous to those in the $\U(1)$ DSL:
\begin{subequations}
    \begin{align}
        &\Delta_{d+id} = e^{-i\frac{\pi}{3}}\left(B_1^\top \kappa^-B_2+B_3^\top \kappa^+B_4\right)\,,\\
        &\Delta_{d-id} = e^{+i\frac{\pi}{3}}\left(B_{1}^\top\kappa^+ B_{2}+B_{3}^\top\kappa^-B_{4}\right)\,,
    \end{align}
\end{subequations}
where $\kappa^{\pm}=1/2(\kappa^1\pm i\kappa^2)$.

Additionally, the non-zero $\mix$ enables an extended $s$-wave pairing:
\begin{equation}
    \Delta_{s^*}=\sin(\mix)\left(B_1^\top \kappa^3 B_2+B_3^\top \kappa^3 B_4\right)\,,
\end{equation}
which vanishes in the $\U(1)$ DSL ($\mix=0$).

The symmetry transformations of these terms are shown in \tabref{tab: transf. SC Z2}.
\begin{table}[b!]
    \centering
    \begin{tblr}{
        width = 0.6 \linewidth,
        colspec = {c  X[c] X[c] X[c]},
        cells = {m},
        rowsep = {3pt}, 
        colsep = {0pt},
    }
    \toprule
    Symmetry & $\Delta_{d+id}$ & $\Delta_{d-id}$ & $\Delta_{s*}$ \\ \midrule
         $T_1$ &$\Delta_{d+id}$ & $\Delta_{d-id}$ & $\Delta_{s^*}$\\
         $T_2$ &$\Delta_{d+id}$ & $\Delta_{d-id}$ & $\Delta_{s^*}$\\
         $\sigma_d$ & $\Delta_{d-id}$ & $\Delta_{d+id}$ & $\Delta_{s^*}$\\
         $C_{6z}$  & $e^{i\frac{2\pi}{3}}\Delta_{d+id}$ & $e^{-i\frac{2\pi}{3}}\Delta_{d-id}$& $\Delta_{s^*}$ \\
         $\mathcal{T}$& $\Delta_{d-id}$ & $\Delta_{d+id}$ & $\Delta_{s^*}$\\\bottomrule
    \end{tblr}
    \caption{Transformation properties of the uniform superconducting order parameters for the $\Z_2$ spin liquid. We reproduce the transformations of the chiral $d\pm id$ states alongside the newly allowed extended $s$-wave pairing for direct comparison.}
    \label{tab: transf. SC Z2}
\end{table}

\subsection{Symmetry-breaking patterns of the chiral spin liquid}
For the chiral spin liquid, as shown in \secref{sec:chargon_minima:CSL}, two of the four low-energy momenta are shifted up in energy, leaving only two valleys; without loss of generality, we take $\phi>0$, so only $\eta=2,3$ contribute. Since these two low-energy modes take the same form as in the U(1) DSL, the corresponding order parameters follow directly from those of the U(1) DSL by removing all terms containing $\eta=1$ or $4$. In the following, we present all order parameters for this case; the symmetry transformations of the surviving fields under $T_1$, $T_2$, $C_{6z}\mathcal{T}$, and $\sigma_d\mathcal{T}$ follow from the ones given in the tables of \secref{sec:U1_orders} by composing the corresponding rows.

\subsubsection{CDW and BDW orders}
In contrast to the $\U(1)$ DSL, the $\rho_{K/2,n}$, $n=1,2,3$ orders vanish, whilst only the CDWs at $M_1$ to $M_3$ persist:
\begin{subequations}
    \begin{align}
        &\rho_{M_1}=\left(B^\dagger_{2}\kappa^3B_{2}-B^\dagger_{3}\kappa^3B_{3}\right)\,,\\
        &\rho_{M_2}=-\left(B^\dagger_{3}\kappa^3B_{2}+B^\dagger_{2}\kappa^3B_{3}\right)\,,\\
        &\rho_{M_3}=i\left(B^\dagger_{3}\kappa^3B_{2}-B^\dagger_{2}\kappa^3B_{3}\right)\,.
    \end{align}
\end{subequations}
Similarly, for the bond density waves, $Q_{K/2,n}$, $n=1,2,3$, vanish:
\begin{subequations}
    \begin{align}
        &Q_{M_1}=\left(B^\dagger_{2}B_{2}-B^\dagger_{3}B_{3}\right)\,,\\
        &Q_{M_2}=-\left(B^\dagger_{3}B_{2}+B^\dagger_{2}B_{3}\right)\,,\\
        &Q_{M_3}=i\left(B^\dagger_{3}B_{2}-B^\dagger_{2}B_{3}\right)\,.
    \end{align}
\end{subequations}

\subsubsection{Superconducting orders}\label{sec:CSL_SC_orders}
The uniform superconducting orders $\Delta_{d\pm id}$ vanish for the CSL, consistent with the observations for the ``Type II'' ansatz studied in Ref.~\cite{Song_2023}.

Similarly, all $M$-point PDWs vanish, and only half of the $K/2$ PDWs can still be formed:
\begin{subequations}
    \begin{align}
        &\Delta_{K/2,-1} = 2B_2^\top\kappa^1 B_3\,,\\
        &\Delta_{K/2,+2} = B_2^\top\kappa^1 B_2-B_3^\top\kappa^1 B_3\,,\\
        &\Delta_{K/2,-3} = -i(B_2^\top\kappa^1 B_2+B_3^\top\kappa^1 B_3)\,.
    \end{align}
\end{subequations}

\subsubsection{Current orders}
The uniform current order is generically induced whenever chargons condense,
\begin{equation}
    J_\Gamma = -|B_2|^2-|B_3|^2\,,
\end{equation}
while the currents $J_{K,E}$ and $J_{K,E^*}$ vanish. All $M$-point current orders are still non-zero:
\begin{subequations}
    \begin{align}
        &J_{M_1,l} = -|B_2|^2+|B_3|^2\,,\\
        &J_{M_2,l} = B^\dagger_2 B_3+B^\dagger_3 B_2\,,\\
        &J_{M_3,l} = i\left(B^\dagger_2 B_3-B^\dagger_3 B_2\right)\,,\\
        &\nonumber\\
        &J_{M_1,s} = -B_2^\dagger\,\kappa^3B_2+B_3^\dagger\,\kappa^3B_3\,,\\
        &J_{M_2,s} = B^\dagger_2\,\kappa^3 B_3+B^\dagger_3 \,\kappa^3B_2\,,\\
        &J_{M_3,s} = i\left(B^\dagger_2\,\kappa^3 B_3-B^\dagger_3\,\kappa^3 B_2\right)\,.
    \end{align}
\end{subequations}

We note that the current orders are identical to the charge orders, up to an arbitrary sign:
\begin{equation}
    J_{M_i,l}=-Q_{M_i}\,, \quad J_{M_i,s}=-\rho_{M_i}\,,\quad i=1,\dots,3\,.
\end{equation}
For the CSL, current and charge/bond orders thus always condense at the same time.

\section{Conclusions and outlook}

We have developed a unified theory of charge fluctuations proximate to
the $\U(1)$ Dirac spin liquid on the triangular lattice and to its
gapped $\mathbb{Z}_2$ and chiral descendants. The charged
degrees of freedom were represented by charge-$e$ bosonic chargons that
carry a fundamental charge under the emergent gauge field. Their
projective symmetry transformations are fixed by those of the fermionic
spinons and by the requirement that their fusion produces the
microscopic electron. This construction allowed us to determine the
chargon band minima and systematically classify the gauge-invariant
orders generated upon chargon condensation. Because condensation of a
fundamental chargon higgses the emergent gauge structure, the resulting
phases have no deconfined gauge excitations, although they can exhibit a
rich pattern of conventional broken symmetries.

For the $\U(1)$ Dirac spin liquid, the chargon spectrum possesses four
symmetry-related minima. Bilinears formed from the corresponding
low-energy fields generate charge- and bond-density waves at the three
$M$ points and at the three $K/2$ wave vectors, uniform
$d\pm id$ superconductivity, several families of pair-density waves,
and uniform or translation-breaking loop-current orders. These orders
are not independent competing fields, but different gauge-invariant
composites of a common fractionalized order parameter. Their coexistence
is therefore an intrinsic consequence of chargon condensation rather
than an accidental coincidence of unrelated instabilities. 

A minimization of a simple short-range lattice functional provides an
illustration of this intertwining. Over the representative parameter
regimes that we studied, the superconducting states are pair-density
waves rather than uniform condensates. Depending on the interactions,
they may preserve or break time reversal and may occur either with a
uniform onsite density or together with a charge-density modulation.
We also find non-superconducting phases in which charge order coexists
with loop currents. Uniform $d\pm id$ superconductivity is allowed by
symmetry but is not selected by this minimal lattice functional. Its
absence in the displayed phase diagrams should therefore be interpreted
as a property of the restricted model and parameter range, and not as a
general obstruction. We emphasize, however, that our analysis was restricted to the mean-field level, and it is possible that gauge field fluctuations suppress density-wave and current-loop orders, stabilizing a pure superconducting state, as shown in a recent study on the square lattice~\cite{Chen2026}.

The gapped descendants of the Dirac spin liquid display characteristic
modifications of this order-parameter content. In the $\mathbb{Z}_2$
spin liquid, mixing between the two chargon gauge components permits
$K$-point charge order and a uniform extended-$s$-wave superconducting
order, both of which vanish continuously upon approaching the $\U(1)$
limit. The $d\pm id$ superconducting channels remain present. In the
chiral spin liquid, by contrast, the emergent flux lifts two of the four
chargon minima. This reduced low-energy manifold eliminates the uniform
superconducting bilinears and the $K/2$ charge and bond orders, while
retaining the $M$-point density waves, a subset of the $K/2$
pair-density waves, and several current orders. The possible ordered
descendants therefore retain information about the projective symmetry
and band structure of their parent spin liquid. Despite the absence of low-energy superconducting bilinears in this chiral spin liquid, superconductivity can still emerge via a different mechanism. As discussed in Ref.~\cite{Song_2023}, the chargons can realize a bosonic integer quantum Hall state, which results in a $d+id$ superconductor~\cite{Song2021}.

Our results provide a symmetry-based framework for interpreting
pressure-, doping-, or light-induced charge dynamics in triangular
lattice spin-liquid candidates. In particular, the simultaneous
appearance of finite-momentum pairing, charge or bond modulations, and
time-reversal-breaking currents would be a natural signature of
the condensation of a fractionalized chargon rather than of a single
conventional order parameter. Conversely, the occurrence of $K$-point
charge order or extended-$s$-wave superconductivity could help
distinguish a proximate $\mathbb{Z}_2$ spin liquid from its $\U(1)$
parent.

An important direction for future work is to understand the dynamics of
the intertwined orders identified here. Charge, bond, current, and
superconducting orders are not independent instabilities, but different
composites of the same chargon field. Their competition and coexistence
near confinement may therefore be substantially altered by chargon and
gauge-field fluctuations, potentially leading to critical behavior that
is not captured by the static Higgs potential. Away from half filling,
the modified chargon dynamics may further change the structure of the
accessible ordered states and the transitions between them. A
particularly promising extension is to driven, potentially light-matter-coupled systems, where the chargon
dispersion and interactions can be modified directly, and selected
ordering tendencies may be enhanced on nonequilibrium time scales. This
could provide a route toward interpreting the light-induced enhancement
of charge coherence observed in triangular-lattice spin-liquid materials~\cite{matteo26}.

\begin{acknowledgements}

We thank Filippo Glerean, Yasir Iqbal, Kazushi Kanoda, Patrick Lee, Atanu Maity, and Matteo Mitrano for valuable discussions. This research was supported by US NSF Grant DMR-2245246. The Flatiron Institute is a division of the Simons Foundation. A.F. and R.T. were supported by the Deutsche Forschungsgemeinschaft (DFG, German Research Foundation) through Project-ID 258499086 -- SFB 1170 and through the W\"urzburg-Dresden Cluster of Excellence on Complexity and Topology in Quantum Matter -- ctd.qmat Project-ID 390858490 -- EXC 2147. 

\end{acknowledgements}

\bibliography{refs.bib}

@ARTICLE{Lu16,
       author = {{Lu}, Yuan-Ming},
        title = "{Symmetric Z$_{2}$ spin liquids and their neighboring phases on triangular lattice}",
      journal = {Phys. Rev. B},
         year = 2016,
        month = apr,
       volume = {93},
       number = {16},
          eid = {165113},
        pages = {165113},
          doi = {10.1103/PhysRevB.93.165113},
archivePrefix = {arXiv},
       eprint = {1505.06495},
 primaryClass = {cond-mat.str-el},
       adsurl = {https://ui.adsabs.harvard.edu/abs/2016PhRvB..93p5113L}
}

@article{Feuerpfeil_2026,
       author = {{Feuerpfeil}, Andreas and {Maity}, Atanu and {Thomale}, Ronny and {Iqbal}, Yasir and {Sachdev}, Subir},
        title = "{Higgs criticality of Dirac spin liquids on depleted triangular lattices}",
      journal = {arXiv e-prints},
         year = 2026,
        month = mar,
          eid = {arXiv:2603.28860},
        pages = {arXiv:2603.28860},
          doi = {10.48550/arXiv.2603.28860},
archivePrefix = {arXiv},
       eprint = {2603.28860},
 primaryClass = {cond-mat.str-el},
       adsurl = {https://ui.adsabs.harvard.edu/abs/2026arXiv260328860F}
}

@ARTICLE{Yasir16,
       author = {{Iqbal}, Yasir and {Hu}, Wen-Jun and {Thomale}, Ronny and {Poilblanc}, Didier and {Becca}, Federico},
        title = "{Spin liquid nature in the Heisenberg $J_{1}$-$J_{2}$ triangular antiferromagnet}",
      journal = {Phys. Rev. B},
         year = 2016,
        month = apr,
       volume = {93},
       number = {14},
          eid = {144411},
        pages = {144411},
          doi = {10.1103/PhysRevB.93.144411},
archivePrefix = {arXiv},
       eprint = {1601.06018},
 primaryClass = {cond-mat.str-el},
       adsurl = {https://ui.adsabs.harvard.edu/abs/2016PhRvB..93n4411I}
}

@ARTICLE{Yasir25,
       author = {{Budaraju}, Sasank and {Parola}, Alberto and {Iqbal}, Yasir and {Becca}, Federico and {Poilblanc}, Didier},
        title = "{Monopole excitations in the U(1) Dirac spin liquid on the triangular lattice}",
      journal = {Phys. Rev. B},
         year = 2025,
        month = mar,
       volume = {111},
       number = {12},
          eid = {125150},
        pages = {125150},
          doi = {10.1103/PhysRevB.111.125150},
archivePrefix = {arXiv},
       eprint = {2410.18747},
 primaryClass = {cond-mat.str-el},
       adsurl = {https://ui.adsabs.harvard.edu/abs/2025PhRvB.111l5150B}
}

@article{Wen2002,
       author = {{Wen}, Xiao-Gang},
        title = "{Quantum orders and symmetric spin liquids}",
      journal = {Phys. Rev. B},
         year = 2002,
        month = apr,
       volume = {65},
       number = {16},
          eid = {165113},
        pages = {165113},
          doi = {10.1103/PhysRevB.65.165113},
archivePrefix = {arXiv},
       eprint = {cond-mat/0107071},
 primaryClass = {cond-mat.str-el},
       adsurl = {https://ui.adsabs.harvard.edu/abs/2002PhRvB..65p5113W}
}

@article{fazekas1974ground,
author = {P. Fazekas and P. W. Anderson},
title = {On the ground state properties of the anisotropic triangular antiferromagnet},
journal = {Philosophical Magazine},
volume = {30},
number = {2},
pages = {423--440},
year = {1974},
publisher = {Taylor \& Francis},
doi = {10.1080/14786439808206568}
}

@ARTICLE{Wietek24,
       author = {{Wietek}, Alexander and {Capponi}, Sylvain and {L{\"a}uchli}, Andreas M.},
        title = "{Quantum Electrodynamics in 2 +1 Dimensions as the Organizing Principle of a Triangular Lattice Antiferromagnet}",
      journal = {Physical Review X},
         year = 2024,
        month = apr,
       volume = {14},
       number = {2},
          eid = {021010},
        pages = {021010},
          doi = {10.1103/PhysRevX.14.021010},
archivePrefix = {arXiv},
       eprint = {2303.01585},
 primaryClass = {cond-mat.str-el},
       adsurl = {https://ui.adsabs.harvard.edu/abs/2024PhRvX..14b1010W}
}

@ARTICLE{Zaletel18,
  title = "{Chiral Spin Liquid Phase of the Triangular Lattice Hubbard Model: A Density Matrix Renormalization Group Study}",
  author = {Szasz, Aaron and Motruk, Johannes and Zaletel, Michael P. and Moore, Joel E.},
  journal = {Phys. Rev. X},
  volume = {10},
  issue = {2},
  pages = {021042},
  numpages = {16},
  year = {2020},
  month = {May},
  publisher = {American Physical Society},
  doi = {10.1103/PhysRevX.10.021042},
archivePrefix = {arXiv},
       eprint = {1808.00463},
 primaryClass = {cond-mat.str-el},
       adsurl = {https://ui.adsabs.harvard.edu/abs/2018arXiv180800463S}
}

@article{Sachdev1992Kagome,
  title = {{Kagome and triangular-lattice Heisenberg antiferromagnets: Ordering from quantum fluctuations and quantum-disordered ground states with unconfined bosonic spinons}},
  author = {Sachdev, Subir},
  journal = {Phys. Rev. B},
  volume = {45},
  issue = {21},
  pages = {12377--12396},
  numpages = {0},
  year = {1992},
  month = {Jun},
  publisher = {American Physical Society},
  doi = {10.1103/PhysRevB.45.12377},
  url = {https://link.aps.org/doi/10.1103/PhysRevB.45.12377}
}

@article{KL87,
  title = "{Equivalence of the resonating-valence-bond and fractional quantum Hall states}",
  author = {Kalmeyer, V. and Laughlin, R. B.},
  journal = {Phys. Rev. Lett.},
  volume = {59},
  issue = {18},
  pages = {2095--2098},
  numpages = {0},
  year = {1987},
  month = {Nov},
  publisher = {American Physical Society},
  doi = {10.1103/PhysRevLett.59.2095},
  url = {https://link.aps.org/doi/10.1103/PhysRevLett.59.2095}
}

@ARTICLE{Jiang23,
       author = {{Jiang}, Yi-Fan and {Jiang}, Hong-Chen},
        title = "{Nature of quantum spin liquids of the S =1/2 Heisenberg antiferromagnet on the triangular lattice: A parallel DMRG study}",
      journal = {Phys. Rev. B},
         year = 2023,
        month = apr,
       volume = {107},
       number = {14},
          eid = {L140411},
        pages = {L140411},
          doi = {10.1103/PhysRevB.107.L140411},
archivePrefix = {arXiv},
       eprint = {2203.10216},
 primaryClass = {cond-mat.str-el},
       adsurl = {https://ui.adsabs.harvard.edu/abs/2023PhRvB.107n0411J}
}

@ARTICLE{Kanoda18,
       author = {{Furukawa}, Tetsuya and {Kobashi}, Kazuhiko and {Kurosaki}, Yosuke and {Miyagawa}, Kazuya and {Kanoda}, Kazushi},
        title = "{Quasi-continuous transition from a Fermi liquid to a spin liquid in {\ensuremath{\kappa}}-(ET)$_{2}$Cu$_{2}$(CN)$_{3}$}",
      journal = {Nature Communications},
         year = 2018,
        month = jan,
       volume = {9},
          eid = {307},
        pages = {307},
          doi = {10.1038/s41467-017-02679-7},
archivePrefix = {arXiv},
       eprint = {1707.05586},
 primaryClass = {cond-mat.str-el},
       adsurl = {https://ui.adsabs.harvard.edu/abs/2018NatCo...9..307F}
}

@article{Kanoda24,
author = {Oike, Hiroshi and Taniguchi, Hiromi and Miyagawa, Kazuya and Kanoda, Kazushi},
title = "{Mottness and Spin Liquidity in a Doped Organic Superconductor $\kappa$-(BEDT-TTF)$_4$Hg$_{2.89}$Br$_8$}",
journal = {Journal of the Physical Society of Japan},
volume = {93},
number = {4},
pages = {042001},
year = {2024},
doi = {10.7566/JPSJ.93.042001}
}

@ARTICLE{Tennant24,
       author = {{Scheie}, A.~O. and {Lee}, Minseong and {Wang}, Kevin and {Laurell}, P. and {Choi}, E.~S. and {Pajerowski}, D. and {Zhang}, Qingming and {Ma}, Jie and {Zhou}, H.~D. and {Lee}, Sangyun and {Huan}, Chao and {Thomas}, S.~M. and {Ajeesh}, M.~O. and {Rosa}, P.~F.~S. and {Chen}, Ao and {Zapf}, Vivien S. and {Heyl}, M. and {Batista}, C.~D. and {Dagotto}, E. and {Moore}, J.~E. and {Tennant}, D. Alan},
        title = "{Spectrum and low-temperature bulk properties of triangular quantum spin liquid candidate NaYbSe$_2$}",
      journal = {arXiv e-prints},
         year = 2024,
        month = jun,
          eid = {arXiv:2406.17773},
        pages = {arXiv:2406.17773},
          doi = {10.48550/arXiv.2406.17773},
archivePrefix = {arXiv},
       eprint = {2406.17773},
 primaryClass = {cond-mat.str-el},
       adsurl = {https://ui.adsabs.harvard.edu/abs/2024arXiv240617773S}
}

@article{chinese_na,
author = {Jia, Ya-Ting  and Gong, Chun-Sheng and Liu, Yi-Xuan and Zhao, Jian-Fa and Dong, Cheng and Dai, Guang-Yang and Li, Xiao-Dong and Lei, He-Chang and Yu, Run-Ze and Zhang, Guang-Ming and Jin, Chang-Qing},
title = "{Mott Transition and Superconductivity in Quantum Spin Liquid Candidate NaYbSe$_2$}",
journal = {Chinese Physics Letters},
volume = {37},
pages = {097404},
year = {2020},
doi = {10.1088/0256-307X/37/9/097404}
}

@article{Song1,
    author = {Song, Xue-Yang and He, Yin-Chen and Vishwanath, Ashvin and Wang, Chong},
    title = "{From spinon band topology to the symmetry quantum numbers of monopoles in Dirac spin liquids}",
    eprint = "1811.11182",
    archivePrefix = "arXiv",
    primaryClass = "cond-mat.str-el",
    doi = "10.1103/PhysRevX.10.011033",
    journal = "Phys. Rev. X",
    volume = "10",
    number = "1",
    pages = "011033",
    year = "2020"
}

@article{matteo26,
    author = {Glerean, Filippo and Bonetti, Pietro M. and Tepie, Meng and Guan, Ziqiang and Edelman, Jaiden and Kelliher, Ty and Priya, Savita and Miyagawa, Kazuya and Kanoda, Kazushi and Dressel, Martin and Sachdev, Subir and Mitrano, Matteo},
    title = "{Light-induced Condensation of a Quantum Spin Liquid}",
    journal = {submitted},
    year = 2026
}

@ARTICLE{Christos23,
       author = {{Christos}, Maine and {Luo}, Zhu-Xi and {Shackleton}, Leyna and {Zhang}, Ya-Hui and {Scheurer}, Mathias S. and {Sachdev}, Subir},
        title = "{A model of $d$-wave superconductivity, antiferromagnetism, and charge order on the square lattice}",
      journal = {Proceedings of the National Academy of Science},
         year = 2023,
        month = may,
       volume = {120},
       number = {21},
          eid = {e2302701120},
        pages = {e2302701120},
          doi = {10.1073/pnas.2302701120},
archivePrefix = {arXiv},
       eprint = {2302.07885},
 primaryClass = {cond-mat.str-el},
       adsurl = {https://ui.adsabs.harvard.edu/abs/2023PNAS..12002701C}
}

@ARTICLE{Bonetti26,
       author = {{Bonetti}, Pietro M. and {Christos}, Maine and {Nikolaenko}, Alexander and {Patel}, Aavishkar A. and {Sachdev}, Subir},
        title = "{Fractionalized Fermi liquids and the cuprate phase diagram}",
      journal = {Reports on Progress in Physics},
         year = 2026,
        month = apr,
       volume = {89},
       number = {4},
          eid = {044501},
        pages = {044501},
          doi = {10.1088/1361-6633/ae530d},
archivePrefix = {arXiv},
       eprint = {2508.20164},
 primaryClass = {cond-mat.str-el},
       adsurl = {https://ui.adsabs.harvard.edu/abs/2026RPPh...89d4501B}
}

@article{Affleck1988,
  title = {{Large-n limit of the Heisenberg-Hubbard model: Implications for high-${T}_{c}$ superconductors}},
  author = {Affleck, Ian and Marston, J. Brad},
  journal = {Phys. Rev. B},
  volume = {37},
  issue = {7},
  pages = {3774--3777},
  numpages = {0},
  year = {1988},
  month = {Mar},
  publisher = {American Physical Society},
  doi = {10.1103/PhysRevB.37.3774},
  url = {https://link.aps.org/doi/10.1103/PhysRevB.37.3774}
}

@article{Song_2023,
	title = "{Deconfined criticalities and dualities between chiral spin liquid, topological superconductor and charge density wave Chern insulator}",
	pages = {215},
	author = {Song, Xue-Yang and Zhang, Ya-Hui},
	journal = {SciPost Phys.},
	volume = {15},
	year = {2023},
	publisher = {SciPost},
	doi = {10.21468/SciPostPhys.15.5.215},
	url = {https://scipost.org/10.21468/SciPostPhys.15.5.215}
}

@article{Zhu-2015,
  title = {{Spin liquid phase of the $S=\frac{1}{2}\phantom{\rule{4.pt}{0ex}}{J}_{1}\ensuremath{-}{J}_{2}$ Heisenberg model on the triangular lattice}},
  author = {Zhu, Zhenyue and White, Steven R.},
  journal = {Phys. Rev. B},
  volume = {92},
  issue = {4},
  pages = {041105},
  numpages = {4},
  year = {2015},
  month = {Jul},
  publisher = {American Physical Society},
  doi = {10.1103/PhysRevB.92.041105},
  url = {https://link.aps.org/doi/10.1103/PhysRevB.92.041105}
}

@article{Hu-2015,
  title = {{Competing spin-liquid states in the spin-$\frac{1}{2}$ Heisenberg model on the triangular lattice}},
  author = {Hu, Wen-Jun and Gong, Shou-Shu and Zhu, Wei and Sheng, D. N.},
  journal = {Phys. Rev. B},
  volume = {92},
  issue = {14},
  pages = {140403},
  numpages = {6},
  year = {2015},
  month = {Oct},
  publisher = {American Physical Society},
  doi = {10.1103/PhysRevB.92.140403},
  url = {https://link.aps.org/doi/10.1103/PhysRevB.92.140403}
}

@misc{Jiang-2026,
      title={{Competing states in the $S=1/2$ triangular-lattice $J_1$-$J_2$ Heisenberg model: a dynamical density-matrix renormalization group study}}, 
      author={Shengtao Jiang and Steven R. White and Steven A. Kivelson and Hong-Chen Jiang},
      year={2026},
      eprint={2602.14892},
      archivePrefix={arXiv},
      primaryClass={cond-mat.str-el},
      url={https://arxiv.org/abs/2602.14892}, 
}

@article{Sherman-2023,
  title = {{Spectral function of the ${J}_{1}\ensuremath{-}{J}_{2}$ Heisenberg model on the triangular lattice}},
  author = {Sherman, Nicholas E. and Dupont, Maxime and Moore, Joel E.},
  journal = {Phys. Rev. B},
  volume = {107},
  issue = {16},
  pages = {165146},
  numpages = {19},
  year = {2023},
  month = {Apr},
  publisher = {American Physical Society},
  doi = {10.1103/PhysRevB.107.165146},
  url = {https://link.aps.org/doi/10.1103/PhysRevB.107.165146}
}

@misc{Drescher-2025,
      title={{Spectral Functions of an Extended Antiferromagnetic $S=1/2$ Heisenberg Model on the Triangular Lattice}}, 
      author={Markus Drescher and Laurens Vanderstraeten and Roderich Moessner and Frank Pollmann},
      year={2025},
      eprint={2508.17292},
      archivePrefix={arXiv},
      primaryClass={cond-mat.str-el},
      url={https://arxiv.org/abs/2508.17292}, 
}

@article{Ferrari-2019,
  title = {{Dynamical Structure Factor of the ${J}_{1}\ensuremath{-}{J}_{2}$ Heisenberg Model on the Triangular Lattice: Magnons, Spinons, and Gauge Fields}},
  author = {Ferrari, Francesco and Becca, Federico},
  journal = {Phys. Rev. X},
  volume = {9},
  issue = {3},
  pages = {031026},
  numpages = {12},
  year = {2019},
  month = {Aug},
  publisher = {American Physical Society},
  doi = {10.1103/PhysRevX.9.031026},
  url = {https://link.aps.org/doi/10.1103/PhysRevX.9.031026}
}

@misc{Kovalska-2026,
      title={{Revisiting the $J_1$-$J_2$ Heisenberg Model on a Triangular Lattice: Quasi-Degenerate Ground States and Phase Competition}}, 
      author={Oleksandra Kovalska and Ester Pagès Fontanella and Benedikt Schneider and Hong-Hao Tu and Jan von Delft},
      year={2026},
      eprint={2603.08650},
      archivePrefix={arXiv},
      primaryClass={cond-mat.str-el},
      url={https://arxiv.org/abs/2603.08650}, 
}

@article{Hu-2019,
  title = {{Dirac Spin Liquid on the Spin-$1/2$ Triangular Heisenberg Antiferromagnet}},
  author = {Hu, Shijie and Zhu, W. and Eggert, Sebastian and He, Yin-Chen},
  journal = {Phys. Rev. Lett.},
  volume = {123},
  issue = {20},
  pages = {207203},
  numpages = {6},
  year = {2019},
  month = {Nov},
  publisher = {American Physical Society},
  doi = {10.1103/PhysRevLett.123.207203},
  url = {https://link.aps.org/doi/10.1103/PhysRevLett.123.207203}
}

@article{Dagotto-1988,
  title = {{SU(2) gauge invariance and order parameters in strongly coupled electronic systems}},
  author = {Dagotto, Elbio and Fradkin, Eduardo and Moreo, Adriana},
  journal = {Phys. Rev. B},
  volume = {38},
  issue = {4},
  pages = {2926--2929},
  numpages = {0},
  year = {1988},
  month = {Aug},
  publisher = {American Physical Society},
  doi = {10.1103/PhysRevB.38.2926},
  url = {https://link.aps.org/doi/10.1103/PhysRevB.38.2926}
}

@ARTICLE{Fierz_1937,
       author = {{Fierz}, Markus},
        title = "{Zur Fermischen Theorie des {\ensuremath{\beta}}-Zerfalls}",
      journal = {Zeitschrift fur Physik},
         year = 1937,
        month = jul,
       volume = {104},
       number = {7-8},
        pages = {553-565},
          doi = {10.1007/BF01330070},
       adsurl = {https://ui.adsabs.harvard.edu/abs/1937ZPhy..104..553F}
}

@ARTICLE{CSS93,
   author = {{Chubukov}, A.~V. and {Senthil}, T. and {Sachdev}, S.},
    title = "{Universal magnetic properties of frustrated quantum antiferromagnets in two dimensions}",
  journal = {Phys. Rev. Lett.},
   eprint = {cond-mat/9311045},
     year = 1994,
    month = mar,
   volume = 72,
    pages = {2089-2092},
      doi = {10.1103/PhysRevLett.72.2089},
   adsurl = {http://adsabs.harvard.edu/abs/1994PhRvL..72.2089C}
}

@article{Wen-1991,
  title = {{Mean-field theory of spin-liquid states with finite energy gap and topological orders}},
  author = {Wen, X. G.},
  journal = {Phys. Rev. B},
  volume = {44},
  issue = {6},
  pages = {2664--2672},
  numpages = {0},
  year = {1991},
  month = {Aug},
  publisher = {American Physical Society},
  doi = {10.1103/PhysRevB.44.2664},
  url = {https://link.aps.org/doi/10.1103/PhysRevB.44.2664}
}

@article{Abrikosov-1965,
  title = {{Electron scattering on magnetic impurities in metals and anomalous resistivity effects}},
  author = {Abrikosov, A. A.},
  journal = {Physics Physique Fizika},
  volume = {2},
  issue = {1},
  pages = {5--20},
  numpages = {16},
  year = {1965},
  month = {Sep},
  publisher = {American Physical Society},
  doi = {10.1103/PhysicsPhysiqueFizika.2.5},
  url = {https://link.aps.org/doi/10.1103/PhysicsPhysiqueFizika.2.5}
}

@article{Bieri_2016,
  title = {Projective symmetry group classification of chiral spin liquids},
  author = {Bieri, Samuel and Lhuillier, Claire and Messio, Laura},
  journal = {Phys. Rev. B},
  volume = {93},
  issue = {9},
  pages = {094437},
  numpages = {28},
  year = {2016},
  month = {Mar},
  publisher = {American Physical Society},
  doi = {10.1103/PhysRevB.93.094437},
  url = {https://link.aps.org/doi/10.1103/PhysRevB.93.094437}
}

@article{Hu_2016,
  title = "{Variational Monte Carlo study of chiral spin liquid in quantum antiferromagnet on the triangular lattice}",
  author = {Hu, Wen-Jun and Gong, Shou-Shu and Sheng, D. N.},
  journal = {Phys. Rev. B},
  volume = {94},
  issue = {7},
  pages = {075131},
  numpages = {7},
  year = {2016},
  month = {Aug},
  publisher = {American Physical Society},
  doi = {10.1103/PhysRevB.94.075131},
  url = {https://link.aps.org/doi/10.1103/PhysRevB.94.075131}
}

@ARTICLE{Chen2026,
       author = {{Chen}, Chuang and {Sachdev}, Subir and {Meng}, Zi Yang},
        title = "{Deconfined criticality between an antiferromagnetic insulator and a nodal d-wave superconductor: a quantum Monte Carlo study}",
      journal = {arXiv e-prints},
         year = 2026,
        month = jul,
          eid = {arXiv:2607.00762},
        pages = {arXiv:2607.00762},
          doi = {10.48550/arXiv.2607.00762},
archivePrefix = {arXiv},
       eprint = {2607.00762},
 primaryClass = {cond-mat.str-el},
       adsurl = {https://ui.adsabs.harvard.edu/abs/2026arXiv260700762C}
}

@ARTICLE{Shi2025,
       author = {{Shi}, Zhengyan Darius and {Senthil}, T.},
        title = "{Doping a Fractional Quantum Anomalous Hall Insulator}",
      journal = {Physical Review X},
         year = 2025,
        month = jul,
       volume = {15},
       number = {3},
          eid = {031069},
        pages = {031069},
          doi = {10.1103/kcm5-hx56},
archivePrefix = {arXiv},
       eprint = {2409.20567},
 primaryClass = {cond-mat.str-el},
       adsurl = {https://ui.adsabs.harvard.edu/abs/2025PhRvX..15c1069S}
}

@ARTICLE{Divic2025,
       author = {{Divic}, Stefan and {Cr{\'e}pel}, Valentin and {Soejima}, Tomohiro and {Song}, Xue-Yang and {Millis}, Andrew J. and {Zaletel}, Michael P. and {Vishwanath}, Ashvin},
        title = "{Anyon superconductivity from topological criticality in a Hofstadter-Hubbard model}",
      journal = {Proceedings of the National Academy of Science},
         year = 2025,
        month = aug,
       volume = {122},
       number = {33},
          eid = {e2426680122},
        pages = {e2426680122},
          doi = {10.1073/pnas.2426680122},
archivePrefix = {arXiv},
       eprint = {2410.18175},
 primaryClass = {cond-mat.str-el},
       adsurl = {https://ui.adsabs.harvard.edu/abs/2025PNAS..12226680D}
}

@ARTICLE{Pichler2026,
       author = {{Pichler}, Fabian and {Kuhlenkamp}, Clemens and {Knap}, Michael and {Vishwanath}, Ashvin},
        title = "{Microscopic Mechanism of Anyon Superconductivity Emerging from Fractional Chern Insulators}",
      journal = {Newton},
      volume = {2},
         year = 2026,
        pages = {100340},
          doi = {10.1016/j.newton.2025.100340},
archivePrefix = {arXiv},
       eprint = {2506.08000},
 primaryClass = {cond-mat.str-el},
       adsurl = {https://ui.adsabs.harvard.edu/abs/2025arXiv250608000P}
}

@ARTICLE{Clemens2025,
       author = {{Kuhlenkamp}, Clemens and {Divic}, Stefan and {Zaletel}, Michael P. and {Soejima}, Tomohiro and {Vishwanath}, Ashvin},
        title = "{Robust superconductivity upon doping chiral spin liquid and Chern insulators in a Hubbard-Hofstadter model}",
      journal = {arXiv e-prints},
         year = 2025,
        month = sep,
          eid = {arXiv:2509.02675},
        pages = {arXiv:2509.02675},
          doi = {10.48550/arXiv.2509.02675},
archivePrefix = {arXiv},
       eprint = {2509.02675},
 primaryClass = {cond-mat.str-el},
       adsurl = {https://ui.adsabs.harvard.edu/abs/2025arXiv250902675K}
}

@ARTICLE{YaHui-ancilla1,
       author = {{Zhang}, Ya-Hui and {Sachdev}, Subir},
        title = "{From the pseudogap metal to the Fermi liquid using ancilla qubits}",
      journal = {Physical Review Research},
         year = 2020,
        month = may,
       volume = {2},
       number = {2},
          eid = {023172},
        pages = {023172},
          doi = {10.1103/PhysRevResearch.2.023172},
archivePrefix = {arXiv},
       eprint = {2001.09159},
 primaryClass = {cond-mat.str-el},
       adsurl = {https://ui.adsabs.harvard.edu/abs/2020PhRvR...2b3172Z}
}

@article{Boulder25,
       author = {{Bonetti}, Pietro M. and {Christos}, Maine and {Patel}, Aavishkar A. and {Sachdev}, Subir},
        title = "{Fractionalized Fermi liquids and the cuprate phase diagram}",
      journal = {Reports on Progress in Physics},
      volume = {89},
      pages = {044501},
         year = {2026},
         doi = {10.1088/1361-6633/ae530d},
archivePrefix = {arXiv},
       eprint = {2508.20164},
 primaryClass = {cond-mat.str-el},
       adsurl = {https://ui.adsabs.harvard.edu/abs/2025arXiv250820164B}
}

@article{Song2021,
  title = {Doping the chiral spin liquid: Topological superconductor or chiral metal},
  author = {Song, Xue-Yang and Vishwanath, Ashvin and Zhang, Ya-Hui},
  journal = {Phys. Rev. B},
  volume = {103},
  issue = {16},
  pages = {165138},
  numpages = {16},
  year = {2021},
  month = {Apr},
  publisher = {American Physical Society},
  doi = {10.1103/PhysRevB.103.165138},
  url = {https://link.aps.org/doi/10.1103/PhysRevB.103.165138}
}

@article{PhysRevLett.86.1881,
  title = "{Resonating Valence Bond Phase in the Triangular Lattice Quantum Dimer Model}",
  author = {Moessner, R. and Sondhi, S. L.},
  journal = {Phys. Rev. Lett.},
  volume = {86},
  issue = {9},
  pages = {1881--1884},
  numpages = {0},
  year = {2001},
  month = {Feb},
  publisher = {American Physical Society},
  doi = {10.1103/PhysRevLett.86.1881},
  url = {https://link.aps.org/doi/10.1103/PhysRevLett.86.1881}
}

@article{PhysRevLett.99.097202,
  title = "{Spin Hamiltonian for which the Chiral Spin Liquid is the Exact Ground State}",
  author = {Schroeter, Darrell F. and Kapit, Eliot and Thomale, Ronny and Greiter, Martin},
  journal = {Phys. Rev. Lett.},
  volume = {99},
  issue = {9},
  pages = {097202},
  numpages = {4},
  year = {2007},
  month = {Aug},
  publisher = {American Physical Society},
  doi = {10.1103/PhysRevLett.99.097202},
  url = {https://link.aps.org/doi/10.1103/PhysRevLett.99.097202}
}

@article{PhysRevB.72.045105,
       author = {{Motrunich}, Olexei I.},
        title = "{Variational study of triangular lattice spin-1/2 model with ring exchanges and spin liquid state in $\kappa$-(ET)$_{2}$Cu$_{2}$(CN)$_{3}$}",
      journal = {Phys. Rev. B},
         year = 2005,
        month = jul,
       volume = {72},
       number = {4},
          eid = {045105},
        pages = {045105},
          doi = {10.1103/PhysRevB.72.045105},
archivePrefix = {arXiv},
       eprint = {cond-mat/0412556},
 primaryClass = {cond-mat.str-el},
       adsurl = {https://ui.adsabs.harvard.edu/abs/2005PhRvB..72d5105M}
}

@article{PhysRevB.91.245125,
  title = "{Phase diagram of the Hubbard model on the anisotropic triangular lattice}",
  author = {Laubach, Manuel and Thomale, Ronny and Platt, Christian and Hanke, Werner and Li, Gang},
  journal = {Phys. Rev. B},
  volume = {91},
  issue = {24},
  pages = {245125},
  numpages = {12},
  year = {2015},
  month = {Jun},
  publisher = {American Physical Society},
  doi = {10.1103/PhysRevB.91.245125},
  url = {https://link.aps.org/doi/10.1103/PhysRevB.91.245125}
}

@article{Takada2003,
	author = {Takada, Kazunori and Sakurai, Hiroya and Takayama-Muromachi, Eiji and Izumi, Fujio and Dilanian, Ruben A. and Sasaki, Takayoshi},
	journal = {Nature},
	number = {6927},
	pages = {53--55},
	title = "{Superconductivity in two-dimensional CoO$_2$ layers}",
	volume = {422},
	year = {2003}}

@article{10.1063/5.0077901,
    author = {Kiese, Dominik and He, Yuchi and Hickey, Ciarán and Rubio, Angel and Kennes, Dante M.},
    title = "{TMDs as a platform for spin liquid physics: A strong coupling study of twisted bilayer WSe$_2$}",
    journal = {APL Materials},
    volume = {10},
    number = {3},
    pages = {031113},
    year = {2022},
    month = {03},
    issn = {2166-532X},
    doi = {10.1063/5.0077901}
}

@article{PhysRevB.83.024402,
  title = {Functional renormalization group for the anisotropic triangular antiferromagnet},
  author = {Reuther, Johannes and Thomale, Ronny},
  journal = {Phys. Rev. B},
  volume = {83},
  issue = {2},
  pages = {024402},
  numpages = {6},
  year = {2011},
  month = {Jan},
  publisher = {American Physical Society},
  doi = {10.1103/PhysRevB.83.024402},
  url = {https://link.aps.org/doi/10.1103/PhysRevB.83.024402}
}

\end{document}